\documentclass[12pt]{article}
\usepackage{amsfonts}
\usepackage{amsmath}
\usepackage{amssymb}
\usepackage{graphicx}
\usepackage{color}
\usepackage[all, knot]{xy}
\usepackage{tikz}

\usepackage{ulem}

\usepackage[utf8]{inputenc}
\usepackage{epstopdf}
\usepackage[footnotesize]{caption}
\usepackage{amsthm}
\usepackage{enumitem}
\usepackage{mathrsfs}

\usepackage[margin=3cm]{geometry}

\def \be {\begin{equation}}
\def \ee {\end{equation}}
\def \bea {\begin{eqnarray}}
\def \eea {\end{eqnarray}}
\def \nn {\nonumber}

\def \rr {\raise.35ex\hbox{\small $\prime$}\kern-.17em{\mbox{\large $\imath$}}}

\def \dels {\partial\kern-.6em /\kern.1em}
\def \As {{A\kern-.5em / \kern.5em}}
\def \Ds {D\kern-.7em / \kern.5em}

\def \ks {k\kern-.5em /}
\def \ls {l\kern-.5em /}

\newcommand{\ci}[1]{}

\newcommand{\ba}{\begin{eqnarray}}
\newcommand{\ea}{\end{eqnarray}}
\newcommand{\bal}{\begin{align}}
\newcommand{\eal}{\end{align}}
\newcommand{\bay}[1]{\left(\begin{array}{#1}}
\newcommand{\eay}{\end{array}\right)}

\newcommand{\hide}[1]{}

\newlist{axioms}{enumerate}{2}
\setlist[axioms,1]{label=\textbf{A\arabic{axiomsi}.}, ref=A\arabic{axiomsi}}
\setlist[axioms,2]{label=\textbf{A\arabic{axiomsi}\rlap{\myEnumCounter{axiomsii}}.},%
                   ref=A\arabic{axiomsi}\myEnumCounter{axiomsii},%
                   align=parleft,%
                   leftmargin=0em,%
                   itemsep=1.4ex,%
                   before={\stepcounter{axiomsi}}}

  \usetikzlibrary{decorations.markings}

\begin{document}

\begin{titlepage}
\begin{center}

\textbf{\LARGE
Non-Locality$\neq$Quantum Entanglement
\vskip.3cm
}
\vskip .5in
{\large
Xingyu Guo$^{a, b}$ \footnote{e-mail address: guoxy@m.scnu.edu.cn}
and Chen-Te Ma$^{a, b, c, d, e}$ \footnote{e-mail address: yefgst@gmail.com}
\\
\vskip 1mm
}
{\sl
$^a$
Guangdong Provincial Key Laboratory of Nuclear Science,\\
 Institute of Quantum Matter,
South China Normal University, Guangzhou 510006, Guangdong, China.
\\
$^b$
Guangdong-Hong Kong Joint Laboratory of Quantum Matter,\\
 Southern Nuclear Science Computing Center, 
South China Normal University, Guangzhou 510006, Guangdong, China.
\\
$^c$ 
Asia Pacific Center for Theoretical Physics, \\
Pohang University of Science and Technology, 
Pohang 37673, Gyeongsangbuk-do, South Korea. 
\\
$^d$
School of Physics and Telecommunication Engineering,\\ 
South China Normal University, Guangzhou 510006, Guangdong, China.
\\
$^e$
The Laboratory for Quantum Gravity and Strings,\\
 Department of Mathematics and Applied Mathematics,\\
University of Cape Town, Private Bag, Rondebosch 7700, South Africa. 
}
\\
\vskip 1mm
\vspace{30pt}
\end{center}
\begin{abstract} 
The unique entanglement measure is concurrence in a 2-qubit pure state. 
The maximum violation of Bell's inequality is monotonically increasing for this quantity. 
Therefore, people expect that pure state entanglement is relevant to the non-locality. 
For justification, we extend the study to three qubits. 
We consider all possible 3-qubit operators with a symmetric permutation. 
When only considering one entanglement measure, the numerical result contradicts expectation. 
Therefore, we conclude ``Non-Locality$\neq$Quantum Entanglement''. 
We propose the generalized $R$-matrix or correlation matrix for the new diagnosis of Quantum Entanglement. 
We then demonstrate the evidence by restoring the monotonically increasing result. 
\end{abstract}
\end{titlepage}

\section{Introduction}
\label{sec:1}
\noindent 
The black-body radiation does not have a proper interpretation from classical physics. 
The experimental results introduce discrete values or quantization to a characterization of objects. 
This surprising observation leads to wave-particle duality and the uncertainty principle. 
People combined all concepts to develop a fundamental theory at an atomic scale, Quantum Mechanics (QM) \cite{Ma:2018efs}. 
The modern description of a particle's motion is not deterministic. 
The complex number and probabilistic interpretation introduce the philosophical problem of QM. 
\\

\noindent
The indeterminism may imply the loss of completeness in QM. 
One naive idea is to introduce hidden variables (describing a more fundamental theory). 
Requiring the independence of separated measurement processes (local realism) can rule out non-physical cases (instantaneous interactions between separate events). 
The locality implies a constraint (Bell's inequality) to correlations of two separated particles \cite{Bell:1964kc}. 
The quantum measurement observed the {\it violation} of Bell's inequality \cite{Clauser:1969ny}. 
At the time, the Bell test experiments still suffered some loopholes without conclusive results. 
Recently, the issues disappeared {\it without} changing the conclusion \cite{Hensen:2015ccp}. 
Hence the fact of violation shows the existence of {\it non-locality}. 
\\

\noindent 
When calculating expectation value of Bell's operator in {\it QM}, ones used two {\it largest} eigenvalues of {\it $R$-matrix} \cite{Verstraete:2002} to show an {\it equivalent} description of maximum violation \cite{Cirelson:1980ry}. 
 The maximum violation is {\it monotonically increasing} with {\it concurrence} \cite{Bennett:1996gf} for {\it all} possible {\it pure} states \cite{Wootters:1997id}. 
 The concurrence is also positively correlated with {\it entanglement entropy}. 
 Hence this result successfully shows that Quantum Entanglement is a {\it necessary} and {\it sufficient} condition of violation for 2-qubit. 
\\

\noindent
{\it Quantum Entanglement} is a phenomenon in which the quantum state of each particle does not have an individual description. 
The dynamics of particles only relies on a set of parameters in Classical Mechanics (CM). 
When {\it Quantum Entanglement} happens, the observation also affects the dynamics. 
Therefore, the parameters of CM are not enough to show a consistent description. 
Hence {\it Quantum Entanglement} should be unique for distinguishing QM and CM. 
Because this phenomenon violates local realism, it prohibits local hidden variable theory. 
\\

\noindent 
For a 2-qubit state, one only has {\it one} choice to perform a partial trace operation. 
Any higher dimensional qubit states have {\it more} than one choice. 
This problem shows the difference between 2-qubit and many-body. 
One main difficulty of many-body Quantum Entanglement is the multi-parameter characterization of Quantum Entanglement. 
One can use the Schmidt decomposition to describe a general 2-qubit pure state by one variable. 
Therefore, the diagnosis of Quantum Entanglement is easy. 
In other words, it is {\it hard} to use a similar way to generalize to a general $n$-qubit state \cite{Chang:2017czx,Chang:2017ygi}. 
Currently, people know the following facts in a 3-qubit state: 
\begin{itemize} 
\item Using the generalized Schmidt decomposition \cite{Peres:1994qv} shows that {\it five} variables are enough for a general 3-qubit state \cite{Acin:2000jx}. 
\item The local operations and classical communication (LOCC) show {\it two} inequivalent entangled classes \cite{Dur:2000zz}. 
\item One {\it cannot} ignore the three-body entanglement measure, 3-tangle, in a general study \cite{Coffman:1999jd}. 
\item A 3-qubit state is realizable in experiments \cite{Aoki:2003,Takeda:2018}. 
\end{itemize}
Therefore, a 3-qubit state contains more than one entanglement measure. 
The genuine tripartite entanglement is a necessary ingredient. 
The progress of techniques provides an opportunity to study many-body Quantum Entanglement in theories and experiments.  
Hence a simple study of exploring the possible generalization of many-body Quantum Entanglement is to show an analytical solution of {\it 3-qubit states}. 
\\

\begin{figure}
\begin{center}
\includegraphics[width=0.5\textwidth]{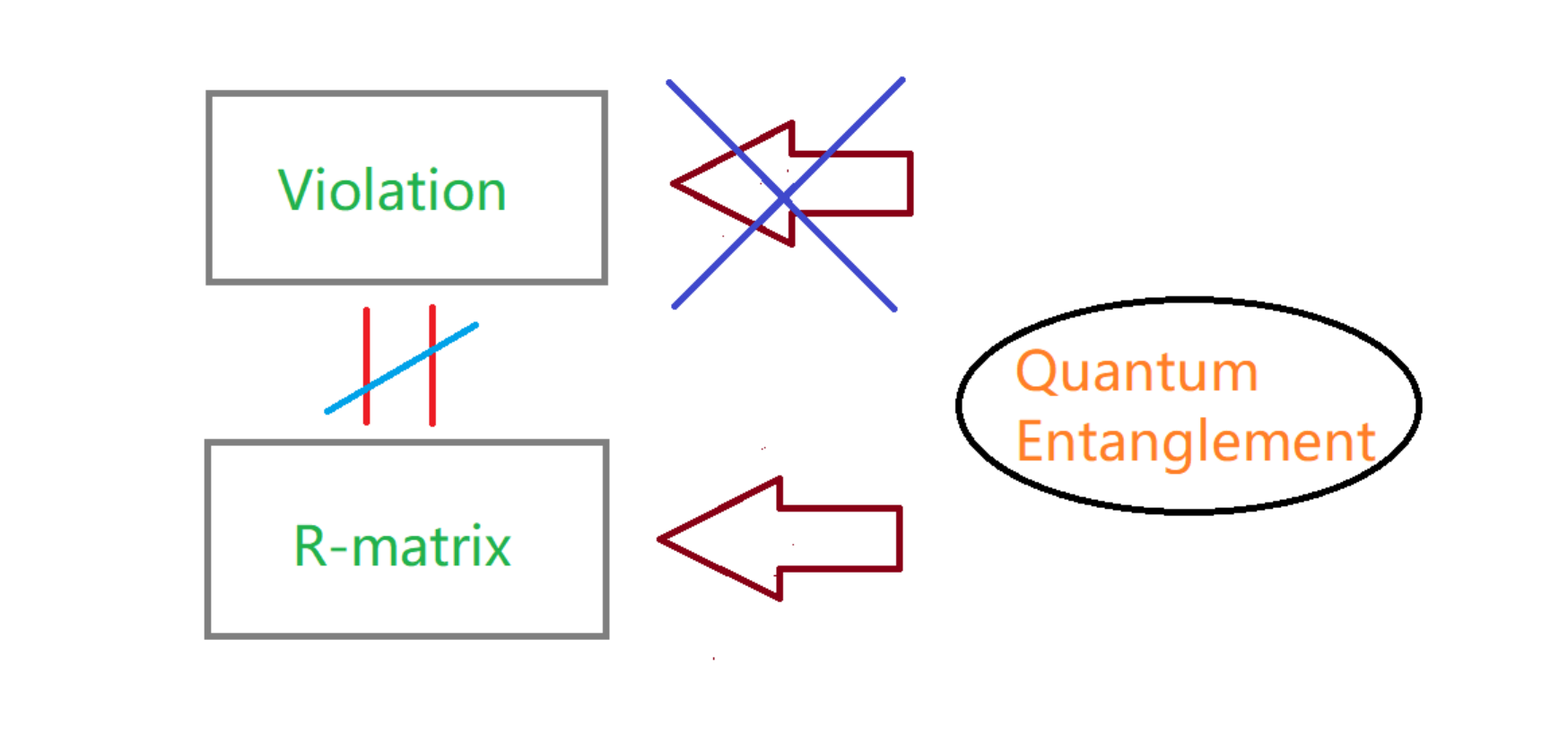}
\end{center}
\caption{We show that ``Violation$\neq$Quantum'' and distinguish the correlation of the $R$-matrix and maximum violation. }
\label{conclusion}
\end{figure}

\noindent
In this paper, we consider all 3-qubit operators with a symmetric permutation. 
Our results justify that Quantum Entanglement is necessary but not sufficient for violation. 
The equivalence in the two-qubit pure state is only a coincidence. 
We then distinguish the maximum violation of Bell's inequality and the correlation of the $R$-matrix. 
The equivalence in two-qubit pure states is again a coincidence. 
We generalize the $R$-matrix and show a diagnosis (Quantum Entanglement). 
We then show our conclusion in Fig. \ref{conclusion}. 
To summarize our results:
\begin{itemize}
\item The characterization of 3-qubit Quantum Entanglement is from five entanglement measures. 
Therefore, it is hard to quantify Quantum Entanglement. 
We discuss turning on one entanglement measure (turning off other measures). 
This case does not have ambiguity for discussing quantification. 
For a proper diagnosis of Quantum Entanglement, monotone behavior must appear. 
We show the loss of monotonically increasing for the maximum violation (consider all possible inequalities). 
Hence it implies ``Non-Locality$\neq$Quantum Entanglement''.  
\item 
In a two-qubit state, the $R$-matrix is 
\bea
R_{i_1i_2}\equiv\mathrm{Tr}(\rho\sigma_{i_1}\otimes\sigma_{i_2}). 
\eea
We consider a naive generalization as the following
\bea
R_{i_1i_2i_3}\equiv\mathrm{Tr}(\rho\sigma_{i_1}\otimes\sigma_{i_2}\otimes\sigma_{i_3}). 
\eea
We then show that the two largest eigenvalues provide the upper bound of maximum violation of Merlin's inequality. 
The analytical solution simultaneously depends on all necessary entanglement measures. 
Therefore, the correlation of the generalized $R$-matrix should generate all 3-qubit Quantum Entanglement.
\item When turning on one entanglement measure, we show a monotonically increasing result from the generalized $R$-matrix. 
Hence this result concretely distinguishes maximum violation from the correlation of the generalized $R$-matrix. 
Since a general 3-qubit state has two different entangled classes, finding a classification \cite{Sawicki:2012,Maciazek:2013,Sawicki:2011,Sawicki:2012(2),Sawicki:2012(3),Maciazek:2017} is unavoidable.  
We realize the classification and show the monotone result for each class. 
\end{itemize}

\noindent
The organization of this paper is as follows: We show ``Non-Locality$\neq$Quantum Entanglement'' by considering all 3-qubit operators in Sec.~\ref{sec:2}. 
We then generalize the $R$-matrix to a 3-qubit state and show that it is a proper diagnosis of Quantum Entanglement \cite{Guo:2021ghs} in Sec.~\ref{sec:3}. 
We discuss our results and conclude in Sec.~\ref{sec:4}. 
We put all numerical results of 3-qubit operators for a single entanglement measure case in \ref{appa}. 
We show the detailed calculation of the generalized $R$-matrix in \ref{appb}. 

\section{Violation$\neq$Quantum}
\label{sec:2}
\noindent 
We first show all possible 3-qubit operators with a symmetric permutation.
Exchanging the qubits does not change the maximum violation. 
We only turn on one entanglement measure for our numerical study. 
The result shows a loss of monotonic relation of maximum violation and the measure. 
Therefore, we show that the maximum degree of violation cannot quantify Quantum Entanglement. 
For convenient reading, we put figures or numerical results in \ref{appa}. 

\subsection{Three-Qubit Operators} 
\noindent 
We construct 3-qubit operators from a linear combination of the following operators:
\bea
&&
A_1\otimes A_2\otimes A_3^{\prime}+A_1\otimes A_2^{\prime}\otimes A_3+A_1^{\prime}\otimes A_2\otimes A_3 ; 
\nn\\
&&
A_1^{\prime}\otimes A_2^{\prime}\otimes A_3+A_1^{\prime}\otimes A_2\otimes A_3^{\prime}+A_1\otimes A_2^{\prime}\otimes A_3 ^{\prime};
\nn\\
&&
A_1^{\prime}\otimes A_2^{\prime}\otimes A_3^{\prime}; 
\nn\\
&&
A_1\otimes A_2\otimes A_3, 
\eea
where 
\bea
A_j\equiv\vec{a}_j\cdot \vec{\sigma}; \qquad A_j^{\prime}\equiv\vec{a^{\prime}}_j\cdot\vec{\sigma}; \qquad 
\vec{\sigma}\equiv(\sigma_x, \sigma_y, \sigma_z).
\eea 
The $\vec{a}$ and $\vec{a^{\prime}}$ are unit vectors:
\bea
\vec{a}\cdot\vec{a}=1; \qquad \vec{a^{\prime}}\cdot\vec{a^{\prime}}=1.
\eea
The notation of the Pauli matrix is given by:
\bea
\sigma_x\equiv
\begin{pmatrix}
0&1
\\
1&0
\end{pmatrix}; \qquad 
\sigma_y\equiv
\begin{pmatrix}
0& -i
\\
i& 0
\end{pmatrix}; \qquad 
\sigma_z\equiv
\begin{pmatrix}
1&0
\\
0&-1
\end{pmatrix}. 
\nn\\
\eea
Each operator is symmetric for exchanging qubits. 
This symmetry also implies invariance of the expectation value of the operators
\bea
\langle{\cal O}\rangle\equiv\mathrm{Tr}(\rho{\cal O}), 
\eea 
where ${\cal O}$ is some operator, and the density matrix is given by
\bea
\rho\equiv|\psi\rangle\langle\psi|. 
\eea
\\

\noindent
 One can observe the maximum violation ($\gamma$) by considering all possible choices of operators (varying $\vec{a}$ and $\vec{a}^{\prime}$)
 \bea
 \gamma\equiv\max_{\cal O}\langle{\cal O}\rangle. 
 \eea
 Hence the maximum violation is invariant under a permutation for the following general 3-qubit operator
\bea
&&
{\cal O}_0
\nn\\
&\equiv&
\bar{\alpha}_1(A_1\otimes A_2\otimes A_3^{\prime}+A_1\otimes A_2^{\prime}\otimes A_3+A_1^{\prime}\otimes A_2\otimes A_3)
\nn\\
&&
+\bar{\alpha}_2(A_1^{\prime}\otimes A_2^{\prime}\otimes A_3+A_1^{\prime}\otimes A_2\otimes  A_3^{\prime}+A_1\otimes A_2^{\prime}\otimes A_3^{\prime})
\nn\\
&&
+\bar{\alpha}_3 A_1^{\prime}\otimes A_2^{\prime}\otimes A_3^{\prime}
\nn\\
&&
+\bar{\alpha}_4 A_1\otimes A_2\otimes A_3, 
\eea 
where 
\bea
-\infty<\bar{\alpha}_1, \bar{\alpha}_2, \bar{\alpha}_3, \bar{\alpha}_4<\infty. 
\eea

\subsection{Three-Qubit State} 
\noindent 
A general 3-qubit state is given by \cite{Acin:2000jx} 
\bea
&&
|\psi\rangle
\nn\\
&=&\lambda_0|000\rangle
+\lambda_1e^{i\phi}|100\rangle
+\lambda_2|101\rangle
+\lambda_3|110\rangle
+\lambda_4|111\rangle, 
\nn\\
&&
\lambda_j\ge 0; \qquad 0\le\phi\le\pi,
\eea 
up to a local unitary transformation. 
Since we normalized the density matrix 
\bea
\mathrm{Tr}\rho=1, 
\eea 
it provides a spherical equation to constrain the coefficients 
\bea
\lambda_0^2+\lambda_1^2+\lambda_2^2+\lambda_3^2+\lambda_4^2=1. 
\eea
Hence a general 3-qubit pure state only has five independent degrees of freedom on the variables. 
Later we will use the quantum state to calculate five necessary entanglement measures. 
Now we show some calculation results. 
\\

\noindent
The density matrix is: 
\bea
&&\rho
\nn\\
&=&
|\psi\rangle\langle\psi|
\nn\\
&=&\lambda_0^2|000\rangle\langle 000|
\nn\\
&&
+\lambda_0\lambda_1e^{-i\phi}|000\rangle\langle 100| 
+\lambda_0\lambda_1e^{i\phi}|100\rangle\langle 000|
\nn\\
&&
+\lambda_0\lambda_2|000\rangle\langle 101|
+\lambda_0\lambda_2|101\rangle\langle 000|
\nn\\
&&
+\lambda_0\lambda_3|000\rangle\langle 110|
+\lambda_0\lambda_3|110\rangle\langle 000|
\nn\\
&&
+\lambda_0\lambda_4|000\rangle\langle 111|
+\lambda_0\lambda_4|111\rangle\langle 000|
+\lambda_1^2|100\rangle\langle 100|
\nn\\
&&
+\lambda_1\lambda_2e^{i\phi}|100\rangle\langle 101|
+\lambda_1\lambda_2e^{-i\phi}|101\rangle\langle 100| 
\nn\\
&&
+\lambda_1\lambda_3e^{i\phi}|100\rangle\langle 110|
+\lambda_1\lambda_3e^{-i\phi}|110\rangle\langle 100|
\nn\\
&&
+\lambda_1\lambda_4e^{i\phi}|100\rangle\langle 111|
+\lambda_1\lambda_4e^{-i\phi}|111\rangle\langle 100| 
\nn\\ 
&&
+\lambda_2^2|101\rangle\langle 101|
\nn\\
&&
+\lambda_2\lambda_3|101\rangle\langle 110|
+\lambda_2\lambda_3|110\rangle\langle 101|
\nn\\ 
&&
+\lambda_2\lambda_4|101\rangle\langle 111|
+\lambda_2\lambda_4|111\rangle\langle 101|
\nn\\
&&
+\lambda_3^2|110\rangle\langle 110|
+\lambda_3\lambda_4|110\rangle\langle 111|
\nn\\ 
&&
+\lambda_3\lambda_4|111\rangle\langle 110|
+\lambda_4^2|111\rangle\langle 111|.
\eea
\\

\noindent
The reduced density matrix of region one is: 
\bea
&&
\rho_1
\nn\\
&=&
\lambda_0^2|0\rangle\langle 0| 
+\lambda_0\lambda_1e^{-i\phi}|0\rangle\langle 1|
+\lambda_0\lambda_1e^{i\phi}|1\rangle\langle 0|
\nn\\
&&
+\lambda_1^2|1\rangle\langle 1|
+\lambda_2^2|1\rangle\langle 1|
+\lambda_3^2|1\rangle\langle 1|
+\lambda_4^2|1\rangle\langle 1|
\nn\\
&=&
\lambda_0^2|0\rangle\langle 0| 
+\lambda_0\lambda_1e^{-i\phi}|0\rangle\langle 1|
+\lambda_0\lambda_1e^{i\phi}|1\rangle\langle 0|
\nn\\
&&
+(1-\lambda_0^2)|1\rangle\langle 1|.
\eea
The reduced density matrix of region two is given by:
\bea
&&
\rho_2
\nn\\
&=&
\lambda_0^2|0\rangle\langle 0|
+\lambda_1^2|0\rangle\langle 0|
+\lambda_1\lambda_3e^{i\phi}|0\rangle\langle 1| 
+\lambda_1\lambda_3e^{-i\phi}|1\rangle\langle 0|
\nn\\
&&
+\lambda_2^2|0\rangle\langle 0|
+\lambda_2\lambda_4|0\rangle\langle 1|
+\lambda_2\lambda_4|1\rangle\langle 0|
\nn\\
&&
+\lambda_3^2|1\rangle\langle 1|
+\lambda_4^2|1\rangle\langle 1|
\nn\\
&=&
(\lambda_0^2+\lambda_1^2+\lambda_2^2)|0\rangle\langle 0|
\nn\\
&&
+(\lambda_2\lambda_4+\lambda_1\lambda_3e^{i\phi})|0\rangle\langle 1|
\nn\\
&&
+(\lambda_2\lambda_4+\lambda_1\lambda_3e^{-i\phi})|1\rangle\langle 0|
\nn\\
&&
+(\lambda_3^2+\lambda_4^2)|1\rangle\langle 1|.
\eea 
The reduced density matrix of region three is given by:
\bea
&&
\rho_3
\nn\\
&=&
\lambda_0^2|0\rangle\langle 0|
+\lambda_1^2|0\rangle\langle 0|
+\lambda_1\lambda_2e^{i\phi}|0\rangle\langle 1|
+\lambda_1\lambda_2e^{-i\phi}|1\rangle\langle 0|
\nn\\
&&
+\lambda_2^2|1\rangle\langle 1|
+\lambda_3^2|0\rangle\langle 0|
+\lambda_3\lambda_4|0\rangle\langle 1|
+\lambda_3\lambda_4|1\rangle\langle 0|
\nn\\
&&
+\lambda_4^2|1\rangle\langle 1|
\nn\\
&=&
(\lambda_0^2+\lambda_1^2+\lambda_3^2)|0\rangle\langle 0|
\nn\\
&&
+(\lambda_3\lambda_4+\lambda_1\lambda_2e^{i\phi})|0\rangle\langle 1|
\nn\\
&&
+(\lambda_3\lambda_4+\lambda_1\lambda_2e^{-i\phi})|1\rangle\langle 0|
\nn\\
&&
+(\lambda_2^2+\lambda_4^2)|1\rangle\langle 1|.
\eea

\subsection{Entanglement Measures}
\noindent
For a 3-qubit quantum state, all invariant quantities under a local unitary transformation are the following:
\bea
I_1&=&\mathrm{Tr}\rho_1^2
\nn\\
&=&\lambda_0^4+2\lambda_0^2\lambda_1^2+(1-\lambda_0^2)^2; 
\nn\\
I_2&=&\mathrm{Tr}\rho_2^2
\nn\\
&=&
(1-\lambda_3^2-\lambda_4^2)^2
+2|\lambda_2\lambda_4+\lambda_1\lambda_3e^{i\phi}|^2
+(\lambda_3^2+\lambda_4^2)^2; 
\nn\\ 
I_3&=&\mathrm{Tr}\rho_3^2
\nn\\
&=&
(1-\lambda_2^2-\lambda_4^2)^2
+2|\lambda_3\lambda_4+\lambda_1\lambda_2e^{i\phi}|^2
+(\lambda_2^2+\lambda_4^2)^2; 
\nn\\
I_4&=&\tau_{1|23}-\tau_{1|2}-\tau_{1|3}; 
\nn\\
I_5&=&\mathrm{Tr}\big((\rho_1\otimes\rho_2)\rho_{12}\big)-\frac{1}{3}\mathrm{Tr}(\rho_1^3)-\frac{1}{3}\mathrm{Tr}(\rho_2^3)
\nn\\
&=&\mathrm{Tr}\big((\rho_2\otimes\rho_3)\rho_{23}\big)-\frac{1}{3}\mathrm{Tr}(\rho_2^3)-\frac{1}{3}\mathrm{Tr}(\rho_3^3)
\nn\\
&=&\mathrm{Tr}\big((\rho_3\otimes\rho_1)\rho_{31}\big)-\frac{1}{3}\mathrm{Tr}(\rho_3^3)-\frac{1}{3}\mathrm{Tr}(\rho_1^3),
\eea
where 
\bea
\tau_{1|23}\equiv 2(1-\mathrm{Tr}\rho_1^2). 
\eea
The $\rho_j$ is a reduced density matrix of the $j$-th qubit. 
The $\sqrt{\tau_{i_1|i_2}}$ is the entanglement of formation of the $i_1$ qubit and $i_2$ qubit after tracing out a qubit \cite{Wootters:1997id}. 
The entanglement of formation is defined by a minimization of $p_j$ and $\psi_j$ as the following \cite{Bennett:1996gf,Wootters:1997id}:
\bea
C(\rho)&\equiv&\min_{p_j, \psi_j}\sum_j p_jC(\psi_j)=\max(0, Q_1-Q_2-Q_3-Q_4), 
\nn\\
&&
Q_1\ge Q_2\ge Q_3\ge Q_4; 
\nn\\
\rho&=&\sum_jp_j|\psi_j\rangle\langle\psi_j|, 
\eea   
where $Q_j$ are the eigenvalues of $\sqrt{\rho(\sigma_y\otimes\sigma_y)\rho^*(\sigma_y\otimes\sigma_y)}$ \cite{Bennett:1996gf,Wootters:1997id}, 
and $C(\psi)$ is the concurrence 
\bea
C(\psi)\equiv\sqrt{2(1-\mathrm{Tr}\rho^2)}. 
\eea
We denote the complex conjugate as $*$. 
The $I_4$ or 3-tangle controls the 3-body entanglement \cite{Coffman:1999jd}. 
The appearance of the 3-body entanglement quantity implies that the 2-body entanglement quantities are not enough \cite{Coffman:1999jd}.  
Now we calculate $I_4$ as in the following: 
\bea
\tau_{1|23}&=&2\big(1-\mathrm{Tr}\rho_1^2\big)
\nn\\
&=&
2\big(1-\lambda_0^4-2\lambda_0^2\lambda_1^2-(1-\lambda_0^2)^2\big);
\eea
\bea
&&
\rho_{12}(\sigma_y\otimes\sigma_y)\rho_{12}^*(\sigma_y\otimes\sigma_y)
\nn\\
&=&
2\lambda_0^3\lambda_3|00\rangle\langle 11|
-\lambda_0^2(2\lambda_1\lambda_3e^{i\phi}+\lambda_2\lambda_4)|00\rangle\langle 01|
\nn\\
&&
+\lambda_0^2(2\lambda_3^2+\lambda_4^2)|00\rangle\langle 00|
\nn\\
&&
+\lambda_0^2(2\lambda_1\lambda_3e^{\i\phi}+\lambda_2\lambda_4)|10\rangle\langle 11|
\nn\\
&&
-2\lambda_0\lambda_1e^{i\phi}(\lambda_1\lambda_3e^{i\phi}+\lambda_2\lambda_4)|10\rangle\langle 01|
\nn\\
&&
+\big(\lambda_0\lambda_1e^{i\phi}(\lambda_3^2+\lambda_4^2)+\lambda_0\lambda_3(\lambda_1\lambda_3e^{i\phi}+\lambda_2\lambda_4)\big)|10\rangle\langle 00| 
\nn\\
&&
+\lambda_0^2(2\lambda_3^2+\lambda_4^2)|11\rangle\langle 11|
\nn\\
&&
-\big(\lambda_0\lambda_3(\lambda_1\lambda_3e^{i\phi}+\lambda_2\lambda_4)+\lambda_0\lambda_1e^{i\phi}(\lambda_3^2+\lambda_4^2)\big)|11\rangle\langle 01| 
\nn\\
&&
+2\lambda_0\lambda_3(\lambda_3^2+\lambda_4^2)|11\rangle\langle 00|;
\nn\\
\eea
\bea
&&
\rho_{13}(\sigma_y\otimes\sigma_y)\rho_{13}^*(\sigma_y\otimes\sigma_y)
\nn\\
&=&
2\lambda_0^3\lambda_2|00\rangle\langle 11|
-\lambda_0^2(2\lambda_1\lambda_2e^{i\phi}+\lambda_3\lambda_4)|00\rangle\langle 01|
\nn\\
&&
+\lambda_0^2(2\lambda_2^2+\lambda_4^2)|00\rangle\langle 00|
\nn\\
&&
+\lambda_0^2(2\lambda_1\lambda_2e^{i\phi}+\lambda_3\lambda_4)|10\rangle\langle 11|
\nn\\
&&
-2\lambda_0\lambda_1e^{i\phi}(\lambda_1\lambda_2e^{i\phi}+\lambda_3\lambda_4)|10\rangle\langle 01|
\nn\\
&&
+\big(\lambda_0\lambda_1e^{i\phi}(\lambda_2^2+\lambda_4^2)+\lambda_0\lambda_2(\lambda_1\lambda_2e^{i\phi}+\lambda_3\lambda_4)\big)|10\rangle\langle 00| 
\nn\\
&&
+\lambda_0^2(2\lambda_2^2+\lambda_4^2)|11\rangle\langle 11|
\nn\\
&&
-\big(\lambda_0\lambda_2(\lambda_1\lambda_2e^{i\phi}+\lambda_3\lambda_4)+\lambda_0\lambda_1e^{i\phi}(\lambda_2^2+\lambda_4^2)\big)|11\rangle\langle 01| 
\nn\\ 
&&
+2\lambda_0\lambda_2(\lambda_2^2+\lambda_4^2)|11\rangle\langle 00|;
\nn\\
\eea
\bea
\tau_{1|23}&=&2\big(1-\lambda_0^4-2\lambda_0^2\lambda_1^2-(1-\lambda_0^2)^2\big)
\nn\\
&=&4\lambda_0^2(1-\lambda_0^2-\lambda_1^2);
\nn\\
\tau_{1|2}&=&4\lambda_0^2\lambda_3^2;
\nn\\
\tau_{1|3}&=&4\lambda_0^2\lambda_2^2.
\eea
Hence we obtain:
\bea
I_4=4\lambda_0^2(1-\lambda_0^2-\lambda_1^2-\lambda_2^2-\lambda_3^2)=4\lambda_0^2\lambda_4^2.
\eea
Here we use the following convenient identities: 
\bea
\sigma_y&=&-i|0\rangle\langle 1|+i|1\rangle\langle 0|; 
\nn\\
\sigma_y\otimes\sigma_y&=&-|00\rangle\langle 11|
+|01\rangle\langle 10|
+|10\rangle\langle 01|
-|11\rangle\langle 00|
\nn\\
\eea
in the calculation. 
\\

\noindent
In the end, we calculate $I_5$ as in the following: 
\bea
&&
\mathrm{Tr}(\rho_1^3)
\nn\\
&=&
\lambda_0^6
+3\lambda_0^2\lambda_1^2
+(1-\lambda_0^2)^3
\nn\\
&=&3\lambda_0^2\lambda_1^2+3\lambda_0^4-3\lambda_0^2+1;
\nn\\
&&
\mathrm{Tr}(\rho_2^3)
\nn\\
&=&(1-\lambda_3^2-\lambda_4^2)^3+3|\lambda_2\lambda_4+\lambda_1\lambda_3e^{i\phi}|^2+(\lambda_3^2+\lambda_4^2)^3;
\nn\\
\eea
\bea
&&
\rho_{12}
\nn\\
&=&
\lambda_0^2|00\rangle\langle 00|
+\lambda_0\lambda_1e^{-i\phi}|00\rangle\langle 10|
+\lambda_0\lambda_1e^{i\phi}|10\rangle\langle 00|
\nn\\
&&
+\lambda_0\lambda_3|00\rangle\langle 11|
+\lambda_0\lambda_3|11\rangle\langle 00|
+\lambda_1^2|10\rangle\langle 10|
\nn\\
&&
\lambda_1\lambda_3e^{i\phi}|10\rangle\langle 11|
+\lambda_1\lambda_3e^{-i\phi}|11\rangle\langle 10|
+\lambda_2^2|10\rangle\langle 10|
\nn\\
&&
+\lambda_2\lambda_4|10\rangle\langle 11|
+\lambda_2\lambda_4|11\rangle\langle 10|
\nn\\
&&
+\lambda_3^2|11\rangle\langle 11|
+\lambda_4^2|11\rangle\langle 11|
\nn\\
&=&
\lambda_0^2|00\rangle\langle 00|
+\lambda_0\lambda_1e^{-i\phi}|00\rangle\langle 10|
+\lambda_0\lambda_1e^{i\phi}|10\rangle\langle 00|
\nn\\
&&
+\lambda_0\lambda_3|00\rangle\langle 11|
+\lambda_0\lambda_3|11\rangle\langle 00|
+(\lambda_1^2+\lambda_2^2)|10\rangle\langle 10|
\nn\\
&&
+(\lambda_1\lambda_3e^{i\phi}+\lambda_2\lambda_4)|10\rangle\langle 11|
\nn\\
&&
+(\lambda_1\lambda_3e^{-i\phi}+\lambda_2\lambda_4)|11\rangle\langle 10|
\nn\\
&&
+(\lambda_3^2+\lambda_4^2)|11\rangle\langle 11|;
\nn\\
&&
\rho_{13}
\nn\\
&=&
\lambda_0^2|00\rangle\langle 00|
+\lambda_0\lambda_1e^{-i\phi}|00\rangle\langle 10|
+\lambda_0\lambda_1e^{i\phi}|10\rangle\langle 00|
\nn\\
&&
+\lambda_0\lambda_2|00\rangle\langle 11|
+\lambda_0\lambda_2|11\rangle\langle 00|
+\lambda_1^2|10\rangle\langle 10|
\nn\\
&&
+\lambda_1\lambda_2e^{i\phi}|10\rangle\langle 11|
+\lambda_1\lambda_2e^{-i\phi}|11\rangle\langle 10|
+\lambda_2^2|11\rangle\langle 11|
\nn\\
&&
+\lambda_3^2|10\rangle\langle 10|
+\lambda_3\lambda_4|10\rangle\langle 11|
+\lambda_3\lambda_4|11\rangle\langle 10|
\nn\\
&&
+\lambda_4^2|11\rangle\langle 11|
\nn\\
&=&
\lambda_0^2|00\rangle\langle 00|
+\lambda_0\lambda_1e^{-i\phi}|00\rangle\langle 10|
+\lambda_0\lambda_1e^{i\phi}|10\rangle\langle 00|
\nn\\
&&
+\lambda_0\lambda_2|00\rangle\langle 11|
+\lambda_0\lambda_2|11\rangle\langle 00|
+(\lambda_1^2+\lambda_3^2)|10\rangle\langle 10|
\nn\\
&&
+(\lambda_1\lambda_2e^{i\phi}+\lambda_3\lambda_4)|10\rangle\langle 11|
\nn\\
&&
+(\lambda_1\lambda_2e^{-i\phi}+\lambda_3\lambda_4)|11\rangle\langle 10| 
\nn\\
&&
+(\lambda_2^2+\lambda_4^2)|11\rangle\langle 11|;
\nn\\
\eea
\bea
&&
\mathrm{Tr}\big((\rho_1\otimes\rho_2)\rho_{12}\big)
\nn\\
&=&
\lambda_0^4+2\lambda_1^2\lambda_0^2+(\lambda_1^2+\lambda_2^2)(1-\lambda_2^2\big)
\nn\\
&&
+\big(-\lambda_0^4+(-\lambda_1^2+\lambda_2^2-\lambda_3^2-\lambda_4^2)\lambda_0^2
\nn\\
&&+(-\lambda_1^2-\lambda_2^2+\lambda_3^2+\lambda_4^2)\big)(\lambda_3^2+\lambda_4^2)
\nn\\
&&
+2|\lambda_1\lambda_3e^{i\phi}+\lambda_2\lambda_4|^2(1-\lambda_0^2)
\nn\\
&&
+\lambda_0^2\lambda_1\lambda_3e^{i\phi}(\lambda_2\lambda_4+\lambda_1\lambda_3e^{-i\phi})
\nn\\
&&
+\lambda_0^2\lambda_1\lambda_3e^{-i\phi}(\lambda_2\lambda_4+\lambda_1\lambda_3e^{i\phi})
\nn\\
&=&
\lambda_0^4+2\lambda_1^2\lambda_0^2+(\lambda_1^2+\lambda_2^2)(1-\lambda_2^2\big)
\nn\\
&&
+\big(2(\lambda_2^2-1)\lambda_0^2-2(\lambda_1^2+\lambda_2^2)+1\big)(\lambda_3^2+\lambda_4^2)
\nn\\
&&
+2|\lambda_1\lambda_3e^{i\phi}+\lambda_2\lambda_4|^2(1-\lambda_0^2)
\nn\\
&&
+\lambda_0^2\lambda_1\lambda_3e^{i\phi}(\lambda_2\lambda_4+\lambda_1\lambda_3e^{-i\phi})
\nn\\
&&
+\lambda_0^2\lambda_1\lambda_3e^{-i\phi}(\lambda_2\lambda_4+\lambda_1\lambda_3e^{i\phi}).
\eea
Therefore, we obtain
\bea
&&
3I_5
\nn\\
&=&
1+3\lambda_0^2(\lambda_0^2-1+\lambda_1^2-\lambda_1^2\lambda_4^2+\lambda_2^2\lambda_3^2)
\nn\\
&&
-3(1-\lambda_0^2)|\lambda_1\lambda_4e^{i\phi}-\lambda_2\lambda_3|^2.
\eea
\\

\noindent 
Now we introduce different invariant quantities (same degrees of freedom as $I_1-I_5$) as in the following:  
\bea
&&
E_1
\nn\\
&\equiv&\tau_{1|2}
\nn\\
&=&2\lambda_0\lambda_3;
\nn\\
&&
E_2
\nn\\
&\equiv&\tau_{1|3}
\nn\\
&=&2\lambda_0\lambda_2; 
\nn\\
&&
E_3
\nn\\
&\equiv&\tau_{2|3}
\nn\\
&=&2|\lambda_1\lambda_4e^{i\phi}-\lambda_2\lambda_3|; 
\nn\\
&&
E_4
\nn\\
&\equiv&\tau
\nn\\
&=&2\lambda_0\lambda_4; 
\nn\\
&&
E_5
\nn\\
&\equiv&\mathrm{Tr}\big((\rho_1\otimes\rho_2)\rho_{12}\big)
-\frac{1}{3}\mathrm{Tr}(\rho_1^3)
-\frac{1}{3}\mathrm{Tr}(\rho_2^3)
\nn\\
&&
+\frac{1}{4}(E_1^2+E_2^2+E_3^2+E_4^2)
\nn\\
&=&\lambda_0^2(\lambda_2^2\lambda_3^2
-\lambda_1^2\lambda_4^2
+|\lambda_1\lambda_4e^{i\phi}-\lambda_2\lambda_3|^2).
\eea
We then can find that the correlation of reduced density matrices is relevant to $E_5$: 
\bea
&&
\mathrm{Tr}\big((\rho_1\otimes\rho_2)\rho_{12}\big)-\mathrm{Tr}(\rho_1^2)-\mathrm{Tr}(\rho_2^2)
\nn\\
&=&
E_5-1+\frac{E_1^2+E_4^2}{4};
\eea
\bea
&&
\mathrm{Tr}\big((\rho_2\otimes\rho_3)\rho_{23}\big)-\mathrm{Tr}(\rho_2^2)-\mathrm{Tr}(\rho_3^2)
\nn\\
&=&
E_5-1-\frac{E_1^2+E_2^2+E_3^2+2E_4^2}{4};
\eea
\bea
&&
\mathrm{Tr}\big((\rho_3\otimes\rho_1)\rho_{31}\big)-\mathrm{Tr}(\rho_3^2)-\mathrm{Tr}(\rho_1^2)
\nn\\
&=&
E_5-1+\frac{E_2^2+E_4^2}{4}.
\eea
Hence the necessity of $I_5$ is due to the correlation of reduced density matrices. 
The invariant quantities $E_1, E_2, E_3, E_4, E_5$ will be helpful in the next section or the generalized $R$-matrix. 

\subsection{Optimization}
\noindent
We do a numerical optimization to obtain the maximum violation. 
In the numerical study, we separate the general case into the following eight operators: 
 \bea
{\cal O}_1\equiv A_1\otimes A_2\otimes A_3^{\prime}+A_1\otimes A_2^{\prime}\otimes A_3+A_1^{\prime}\otimes A_2\otimes A_3 ; 
\nn\\
\eea
\bea
&&
{\cal O}_2
\nn\\
&\equiv&|\tilde{\alpha}_1|(A_1\otimes A_2\otimes A_3^{\prime}+A_1\otimes A_2^{\prime}\otimes A_3+A_1^{\prime}\otimes A_2\otimes A_3) 
\nn\\
&&
+|\tilde{\alpha}_2|(A_1^{\prime}\otimes A_2^{\prime}\otimes A_3+A_1^{\prime}\otimes A_2\otimes A_3^{\prime}
\nn\\
&&
+A_1\otimes A_2^{\prime}\otimes A_3 ^{\prime}); 
\eea
\bea
&&
{\cal O}_3
\nn\\
&\equiv&
\tilde{\alpha}_1(A_1\otimes A_2\otimes A_3^{\prime}+A_1\otimes A_2^{\prime}\otimes A_3+A_1^{\prime}\otimes A_2\otimes A_3) 
\nn\\
&&
+\tilde{\alpha}_2A_1^{\prime}\otimes A_2^{\prime}\otimes A_3^{\prime};
\eea
\bea
&&
{\cal O}_4
\nn\\
&\equiv&
|\tilde{\alpha}_1|(A_1\otimes A_2\otimes A_3^{\prime}+A_1\otimes A_2^{\prime}\otimes A_3+A_1^{\prime}\otimes A_2\otimes A_3) 
\nn\\
&&
+|\tilde{\alpha}_2|A_1\otimes A_2\otimes A_3; 
\eea
\bea
{\cal O}_5\equiv
\tilde{\alpha}_1A_1^{\prime}\otimes A_2^{\prime}\otimes A_3^{\prime} 
+\tilde{\alpha}_2A_1\otimes A_2\otimes A_3; 
\eea
\bea
&&
{\cal O}_6
\nn\\
&\equiv&
\tilde{\alpha}_1(A_1\otimes A_2\otimes A_3^{\prime}+A_1\otimes A_2^{\prime}\otimes A_3+A_1^{\prime}\otimes A_2\otimes A_3)
\nn\\
&&
+\tilde{\alpha}_2(A_1^{\prime}\otimes A_2^{\prime}\otimes A_3+A_1^{\prime}\otimes A_2\otimes  A_3^{\prime}+A_1\otimes A_2^{\prime}\otimes A_3^{\prime})
\nn\\
&&
+\tilde{\alpha}_3 A_1^{\prime}\otimes A_2^{\prime}\otimes A_3^{\prime};
\eea
\bea
&&
{\cal O}_7
\nn\\
&\equiv&
\tilde{\alpha}_1(A_1\otimes A_2\otimes A_3^{\prime}+A_1\otimes A_2^{\prime}\otimes A_3+A_1^{\prime}\otimes A_2\otimes A_3)
\nn\\
&&
+\tilde{\alpha}_2 A_1^{\prime}\otimes A_2^{\prime}\otimes A_3^{\prime}
\nn\\
&&
+\tilde{\alpha}_3 A_1\otimes A_2\otimes A_3; 
\eea
\bea
&&
{\cal O}_8
\nn\\
&\equiv&
\tilde{\alpha}_1(A_1\otimes A_2\otimes A_3^{\prime}+A_1\otimes A_2^{\prime}\otimes A_3+A_1^{\prime}\otimes A_2\otimes A_3)
\nn\\
&&
+\tilde{\alpha}_2(A_1^{\prime}\otimes A_2^{\prime}\otimes A_3+A_1^{\prime}\otimes A_2\otimes  A_3^{\prime}+A_1\otimes A_2^{\prime}\otimes A_3^{\prime})
\nn\\
&&
+\tilde{\alpha}_3 A_1^{\prime}\otimes A_2^{\prime}\otimes A_3^{\prime}
\nn\\
&&
+\tilde{\alpha}_4 A_1\otimes A_2\otimes A_3. 
\eea 
Here we consider the non-zero coefficients 
\bea
0<|\tilde{\alpha}_1|, |\tilde{\alpha}_2|, |\tilde{\alpha}_3|, |\tilde{\alpha}_4|<\infty.
\eea
\\

\noindent
We do not have a mixed term of $A_j$ and $A_j^{\prime}$ in ${\cal O}_5$. 
Therefore, it is easy to show that 
\bea
\gamma\propto \tilde{\alpha}_1+\tilde{\alpha}_2. 
\eea 
The choice of coefficients does not change the conclusion in ${\cal O}_5$. 
\\

\noindent 
Without an ambiguity of interpretation, we only turn on one entanglement measure. 
The entanglement diagnosis must be monotonic increasing for the measure. 
Now we discuss the one entanglement measure. 
Turning off $\lambda_2$ and $\lambda_4$ provides the only non-vanishing $E_1$. 
When turning off $\lambda_3$ and $\lambda_4$, the only non-vanishing measure is $E_2$. 
For the case of $E_3$, one only needs to turn off $\lambda_0$. 
In the end, we choose: 
\bea
\lambda_1=\lambda_2=\lambda_3=0
\eea 
to leave the only non-vanishing $E_4$ or 3-tangle. 
\\

\noindent 
Now we study the numerical solution for $\langle {\cal O}_j\rangle$ for the single measure case. 
For a convenient reading of the main context, we put the numerical results or figures in \ref{appa}. 
For a proper presentation, we present our result for a part of the $\tilde{\alpha}_j$ parameter space. 
Our physical conclusion and result presented also hold for other parameter spaces. 
Because all operators are symmetric in the permutation of the three qubits, the result of $\langle {\cal O}_j\rangle$ has the redundant behavior for $E_1$, $E_2$, and $E_3$. 
One can observe the above phenomenon in Figs. \ref{2-1}, \ref{2-2}, \ref{2-3}, \ref{2-4}, and \ref{2-5}. 
Without showing too much same information, we only calculate $E_1^2$ and $E_4^2$ for $\langle{\cal O}_6\rangle$, $\langle{\cal O}_7\rangle$, and $\langle{\cal O}_8\rangle$ in Figs. \ref{3-1}, \ref{3-2}, and \ref{4}. 
Because all results show the loss of monotonically increasing, we conclude that the non-locality is not equivalent to Quantum Entanglement. 

\section{Generalized $R$-Matrix}
\label{sec:3}
\noindent 
We introduce an alternative diagnosis, the generalized $R$-matrix. 
We then show the monotonic result for one entanglement measure. 
The analytical solution generates one classification of all 3-qubit quantum states. 
In each class, the monotonically increasing result also holds.  
The details of the generalized $R$-matrix is in \ref{appb}. 

\subsection{Generalized $R$-Matrix and Merlin's Operator} 
\noindent
The Merlin's operator ${\cal M}$ is ${\cal O}_3$ with the choice of coefficients: 
\bea
\tilde{\alpha}_1=-\tilde{\alpha}_2=1. 
\eea
We can rewrite the expectation value of ${\cal M}$ in terms of the generalized $R$-matrix: 
\bea
&&
\langle {\cal M}\rangle
\nn\\
&=&
\sum_{i_1, i_2, i_3}
\bigg(
a_{1, i_1}a_{2, i_2}a_{3, i_3}^{\prime}
+a_{1, i_1}a_{2, i_2}^{\prime}a_{3, i_3}
\nn\\
&&
+a_{1, i_1}^{\prime}a_{2, i_2}a_{3, i_3}
-a_{1, i_1}^{\prime}a_{2, i_2}^{\prime}a_{3, i_3}^{\prime}
\bigg)
\nn\\
&&\times 
R_{i_1i_2i_3}
\nn\\
&=&
\bigg(a_1, a_2^TRa_3^{\prime}\bigg)
+\bigg(a_1, a_2^{\prime T}Ra_3\bigg)
\nn\\
&&
+\bigg(a_1^{\prime}, a_2^TRa_3\bigg)
-\bigg(a_1^{\prime}, a_2^{\prime T}Ra_3^{\prime}\bigg),
\eea
where 
\bea
&&
a_j\equiv
\begin{pmatrix}
a_{j, x}
\\
a_{j, y}
\\
a_{j, z}
\end{pmatrix}; \qquad 
a^{\prime}_j\equiv
\begin{pmatrix}
a^{\prime}_{j, x}
\\
a^{\prime}_{j, y}
\\
a^{\prime}_{j, z}
\end{pmatrix}, 
\nn\\ 
&&
R_{i_1i_2i_3}\equiv\mathrm{Tr}(\rho\sigma_{i_1}\otimes\sigma_{i_2}\otimes\sigma_{i_3}).
\eea
We indicate a transpose operation as the superscript $T$. 
The generalized $R$-matrix is given by:
\bea
R&\equiv&(R_x, R_y, R_z), 
\nn\\
R_x&\equiv&
\begin{pmatrix}
R_{xxx}& R_{xxy}& R_{xxz}
\\
R_{xyx}& R_{xyy} & R_{xyz}
\\
R_{xzx}& R_{xzy} & R_{xzz}
\end{pmatrix}; 
\nn\\
R_y&\equiv&
\begin{pmatrix}
R_{yxx}& R_{yxy}& R_{yxz}
\\
R_{yyx}& R_{yyy} & R_{yyz}
\\
R_{yzx}& R_{yzy} & R_{yzz}
\end{pmatrix}; 
\nn\\
R_z&\equiv&
\begin{pmatrix}
R_{zxx}& R_{zxy}& R_{zxz}
\\
R_{zyx}& R_{zyy} & R_{zyz}
\\
R_{zzx}& R_{zzy} & R_{zzz}
\end{pmatrix}.
\eea
We define the inner product as: 
\bea
&&
\bigg(a_1, a_2^T\vec{R}a_3^{\prime}\bigg)
\nn\\
&\equiv&
\Bigg( \bigg(a_1, a_2^T R_x a_3^{\prime}\bigg), \bigg(a_1, a_2^T R_y a_3^{\prime}\bigg), \bigg(a_1, a_2^T R_x a_3^{\prime}\bigg)\Bigg)
\nn\\
&\equiv& 
\sum_{i_1, i_2, i_3} a_{1, i_1}a_{2, i_2}a_{3, i_3}^{\prime}R_{i_1, i_2, i_3}.
\eea 
\\ 

\noindent 
We show that the generalized $R$-matrix can provide an upper bound to $\langle{\cal M}\rangle$. 
We first observe that the following vectors are orthogonal: 
\bea
&&
V\equiv V_{j, k}=
 \begin{pmatrix}
a_{2, j}a_{3, k}^{\prime}+a_{2, j}^{\prime}a_{3, k}
\end{pmatrix}; 
\nn\\
&&
V^{\prime}\equiv V^{\prime}_{j, k}=
\begin{pmatrix}
a_{2, j}a_{3, k}-a_{2, j}^{\prime}a_{3, k}^{\prime}
\end{pmatrix}, 
\nn\\
&&
\sum_{j ,k=1}^3 V_{j, k}V^{\prime}_{j, k}=0. 
\eea
The norm of the two vectors is:
\bea
|V|^2\equiv V_{j, k}V_{j, k}
&=&
2+2\cos(\theta_2)\cos(\theta_3); 
\nn\\ 
 |V^{\prime}|^2\equiv V^{\prime}_{j, k}V^{\prime}_{j, k} 
 &=&
 2-2\cos(\theta_2)\cos(\theta_3),
\eea
where 
\bea
&&
\vec{a}_2\cdot\vec{a^{\prime}}_2\equiv\cos(\theta_2); \qquad 
\vec{a}_3\cdot\vec{a^{\prime}}_3\equiv\cos(\theta_3);  
\nn\\
&&
0\le\theta_2, \theta_3\le \pi.
\eea 
We then introduce the orthogonal unit vectors ($c$ and $c^{\prime}$) as in the following: 
\bea
V\equiv 2c\cos(\theta); \qquad V^{\prime}\equiv 2c^{\prime}\sin(\theta), 
\eea
where 
\bea
\cos(2\theta)\equiv\cos(\theta_2)\cos(\theta_3),\ 0\le\theta\le\frac{\pi}{2}.
\eea
Therefore, $\langle{\cal M}\rangle$ becomes 
\bea
\langle {\cal M}\rangle=
2\cos(\theta)\big( a_1, R c\big)+2 \sin(\theta)\big(a_1^{\prime}, R c^{\prime}\big).
\eea
Because $c$ and $c^{\prime}$ are not independent, we only obtain an upper bound of maximum violation: 
\bea
\gamma\le 2\sqrt{u_1^2+u_2^2}, 
\eea 
where $u_1^2$ and $u_2^2$ are two largest eigenvalues of $RR^T$. 
The generalized $R$-matrix now has one 3d index and one 9d index. 
Therefore, we can have three possible choices: 
\bea
R^{(1)}_{j_1J_1}&\equiv& R_{j_1j_2j_3}|_{J_1=(j_2, j_3)}; 
\nn\\ 
 R^{(2)}_{j_2J_2}&\equiv& R_{j_1j_2j_3}|_{J_2=(j_1, j_3)}; 
 \nn\\ 
  R^{(3)}_{j_3J_3}&\equiv& R_{j_1j_2j_3}|_{J_3=(j_1, j_2)}, 
\eea
where $j_1, j_2, j_3=x, y, z$. 
To obtain a tight bound of maximum violation, we define a new quantity $\gamma_R$ as that: 
\bea
\gamma\le\gamma_R=2\min_{R^{(1)}, R^{(2)}, R^{(3)}}\sqrt{u_1^2+u_2^2}. 
\eea
Later we will rewrite $\gamma_R$ from five entanglement quantities ($E_{1, 2, 3, ,4 ,5}$). 
This result implies that 3-qubit Quantum Entanglement is encoded by $\gamma_R$.  

\subsection{Eigenvalues of Generalized $R$-Matrix} 
\noindent 
We solve the eigenvalues ($x^{(j)}$) of 
\bea
R^{(j)}R^{(j)T}\equiv M^{(j)}
\eea
from the following equation
\bea
&&
x^{(j)3}
+(-M^{(j)}_{xx}-M^{(j)}_{yy}-M^{(j)}_{zz})x^{(j)2}
\nn\\
&&
+(M^{(j)}_{xx}M^{(j)}_{yy}+M^{(j)}_{xx}M^{(j)}_{zz}+M^{(j)}_{yy}M^{(j)}_{zz}
\nn\\
&&
-M_{xy}^{(j)2}-M_{xz}^{(j)2}-M_{yz}^{(j)2})x^{(j)}
\nn\\
&&
+(-M_{xx}^{(j)}M_{yy}^{(j)}M_{zz}^{(j)}
\nn\\
&&
+M_{xx}^{(j)}M_{yz}^{(j)2}+M_{yy}^{(j)}M_{xz}^{(j)2}+M_{zz}^{(j)}M_{xy}^{(j)2}
\nn\\
&&
-2M_{xy}^{(j)}M_{yz}^{(j)}M_{xz}^{(j)})
\nn\\
&&
=
0.
\eea
Therefore, we can obtain an analytical solution by solving the cubic equation.
Because the eigenvalues are real-valued, the discriminant is non-positive
\bea
&&
\Delta^{(j)}
\nn\\
&\equiv&\bigg(-\frac{\big(\alpha_1^{(j)}\big)^{3}}{27}-\frac{\alpha_3^{(j)}}{2}+\frac{\alpha_1^{(j)}\alpha_2^{(j)}}{6}\bigg)^2
\nn\\
&&
+\bigg(\frac{\alpha_2^{(j)}}{3}-\frac{\big(\alpha_1^{(j)}\big)^2}{9}\bigg)^3\le 0, 
\eea
where 
\bea
\gamma_1^{(j)}&\equiv&-\frac{\alpha_1^{(j)3}}{27}-\frac{\alpha^{(j)}_3}{2}+\frac{\alpha^{(j)}_1\alpha^{(j)}_2}{6}; 
\nn\\ 
\gamma_2^{(j)}&\equiv& \frac{\alpha_2^{(j)}}{3}-\frac{\alpha_1^{(j)2}}{9}\le 0, 
\eea
\bea
&&
\alpha^{(j)}_1
\nn\\
&=&-M^{(j)}_{xx}-M^{(j)}_{yy}-M^{(j)}_{zz}\le 0; 
\nn\\
&&
\alpha^{(j)}_2
\nn\\
&=&M^{(j)}_{xx}M^{(j)}_{yy}+M^{(j)}_{xx}M^{(j)}_{zz}+M^{(j)}_{yy}M^{(j)}_{zz}
\nn\\
&&
-M_{xy}^{(j)2}-M_{xz}^{(j)2}-M_{yz}^{(j)2}; 
\nn\\
&&
\alpha_3^{(j)}
\nn\\
&=&-M_{xx}^{(j)}M_{yy}^{(j)}M_{zz}^{(j)}
\nn\\
&&
+M_{xx}^{(j)}M_{yz}^{(j)2}+M_{yy}^{(j)}M_{xz}^{(j)2}+M_{zz}^{(j)}M_{xy}^{(j)2}
\nn\\
&&
-2M_{xy}^{(j)}M_{yz}^{(j)}M_{xz}^{(j)}.
\eea
The analytical solution of eigenvalues is: 
\bea
&&
x_1^{(j)}
\nn\\
&=&-\frac{\alpha_1^{(j)}}{3}
\nn\\
&&
+2\sqrt{-\gamma^{(j)}_2}\cos\bigg\lbrack
\frac{1}{3}\arccos\bigg(\frac{\gamma_1^{(j)}}{(-\gamma_2^{(j)})^{\frac{3}{2}}}\bigg)\bigg\rbrack; 
\nn\\
&&
x_2^{(j)}
\nn\\
&=&-\frac{\alpha_1^{(j)}}{3}
\nn\\
&&
+2\sqrt{-\gamma^{(j)}_2}\cos\bigg\lbrack
\frac{1}{3}\arccos\bigg(\frac{\gamma_1^{(j)}}{(-\gamma_2^{(j)})^{\frac{3}{2}}}\bigg)+\frac{2\pi}{3}\bigg\rbrack; 
\nn\\
&&
x_3^{(j)}
\nn\\
&=&-\frac{\alpha_1^{(j)}}{3}
\nn\\
&&
+2\sqrt{-\gamma^{(j)}_2}\cos\bigg\lbrack
\frac{1}{3}\arccos\bigg(\frac{\gamma_1^{(j)}}{(-\gamma_2^{(j)})^{\frac{3}{2}}}\bigg)-\frac{2\pi}{3}\bigg\rbrack.
\eea
\\

\noindent 
Now we use the details of \ref{appb}
to rewrite $\alpha_{1}^{(1)}$, $\alpha_{2}^{(1)}$, and $\alpha_{3}^{(1)}$ in terms of entanglement quantities: 
\bea
&&
\alpha_1^{(1)}
\nn\\
&=&-1-(2E_1^2+2E_2^2+2E_3^2+3E_4^2)
\nn\\
&=&-1-(C_1^2+C_2^2+C_3^2)
\nn\\
&\equiv&-1-C_T^2; 
\nn\\
&&
\alpha_2^{(1)}
\nn\\
&=&2(E_1^2+E_2^2+E_4^2)E_3^2+2(E_1^2+E_2^2)(E_4^2+1)
\nn\\
&&
+E_1^4+E_2^4+4E_4^2+16E_5;
\nn\\
&&
\alpha_3^{(1)}
\nn\\
&=&(E_1^2+E_2^2+2E_3^2+2E_4^2)
\nn\\
&&\times
(2E_4^4+2E_1^2E_2^2+E_1^2E_4^2+E_2^2E_4^2)
\nn\\
&&
-(E_1^2+E_2^2+2E_4^2+8E_5)^2.
\eea
The non-negative total concurrence 
\bea
C_T^2=C_1^2+C_2^2+C_3^2, 
\eea
where 
\bea
C_1(\psi)&\equiv&\sqrt{2(1-\mathrm{Tr}\rho_1)}=\sqrt{E_1^2+E_2^2+E_4^2}; 
\nn\\
C_2(\psi)&\equiv&\sqrt{2(1-\mathrm{Tr}\rho_2)}=\sqrt{E_1^2+E_3^2+E_4^2}; 
\nn\\ 
C_3(\psi)&\equiv&\sqrt{2(1-\mathrm{Tr}\rho_3)}=\sqrt{E_2^2+E_3^2+E_4^2}, 
\eea
implies that 
\bea
\alpha_1^{(1)}<0. 
\eea
For $\alpha_2^{(1)}$, the only negative contribution, $-\lambda_0^2\lambda_1^2\lambda_4^2$ is in $E_5$. 
We can combine $4E_4^2$ with $16E_5$ to cancel the negative contribution as that:
\bea
4E_4^2-16\lambda_0^2\lambda_1^2\lambda_4^2
&=&16(\lambda_0^2\lambda_4^2-\lambda_0^2\lambda_1^2\lambda_4^2)
\nn\\
&=&16\lambda_0^2\lambda_4^2(1-\lambda_1^2)\ge 0.
\eea
Hence $\alpha_2^{(1)}$ is not negative. 
We can use the following exchange to obtain other cases: 
\bea
&&
E_2\longleftrightarrow E_3, \qquad 
\alpha_1^{(1)}\leftrightarrow\alpha_1^{(2)}, \alpha_2^{(1)}\leftrightarrow\alpha_2^{(2)}, \alpha_3^{(1)}\leftrightarrow\alpha_3^{(2)}; 
\nn\\
&&
E_1\longleftrightarrow E_3, \qquad 
\alpha_1^{(1)}\leftrightarrow\alpha_1^{(3)}, \alpha_2^{(1)}\leftrightarrow\alpha_2^{(3)}, \alpha_3^{(1)}\leftrightarrow\alpha_3^{(3)}.  
\nn\\
\eea
Because $E_4$ is invariant for a different choice of generalized $R$-matrix, $\alpha_1^{(j)}$ is independent of the index $j$. 
One non-trivial fact is that $E_5$ is also invariant because it depends on $E_{1, 2, 3}$. 
Therefore, using $E_5$ is more convenient than $I_5$. 
Due to the invariance property of $E_4$ and $E_5$, we can show that
\bea
\alpha_2^{(2)}, \alpha_2^{(3)}\ge 0.
\eea 
The eigenvalues of $RR^T$ are functions of $\alpha_{1, 2, 3}$. 
Therefore, it implies that 3-qubit entanglement information is all in $\gamma_R$. 
\\

\noindent 
Now we show an analytical solution of $\gamma_R$. 
Indeed, we know that $x_2^{(j)}$ is always negative, $x_1^{(j)}$ is always positive, and
\bea
x_3^{(j)}\ge x_2^{(j)}, 
\eea
which is due to the following ranges:
  \bea
0\le\theta^{(j)}\equiv\frac{1}{3}\arccos\bigg(\frac{\gamma^{(j)}_1}{(-\gamma^{(j)}_2)^{\frac{3}{2}}}\bigg)\le\frac{\pi}{3}.
\eea
Therefore, two largest eigenvalues of $R^{(j)}R^{(j)T}$ are $x_1^{(j)}$ and $x_3^{(j)}$. 
Indeed, one can also show that the maximum eigenvalue is $x_1^{(j)}$. 
Hence the analytical solution is
\bea
\gamma_R=2\min_j\sqrt{-\frac{2\alpha_1^{(j)}}{3}+2\sqrt{-\gamma^{(j)}_2}\cos\bigg(\theta^{(j)}-\frac{\pi}{3}\bigg)}.
\nn\\
\eea  
\\

\begin{figure}
\begin{center}
\includegraphics[width=0.5\textwidth]{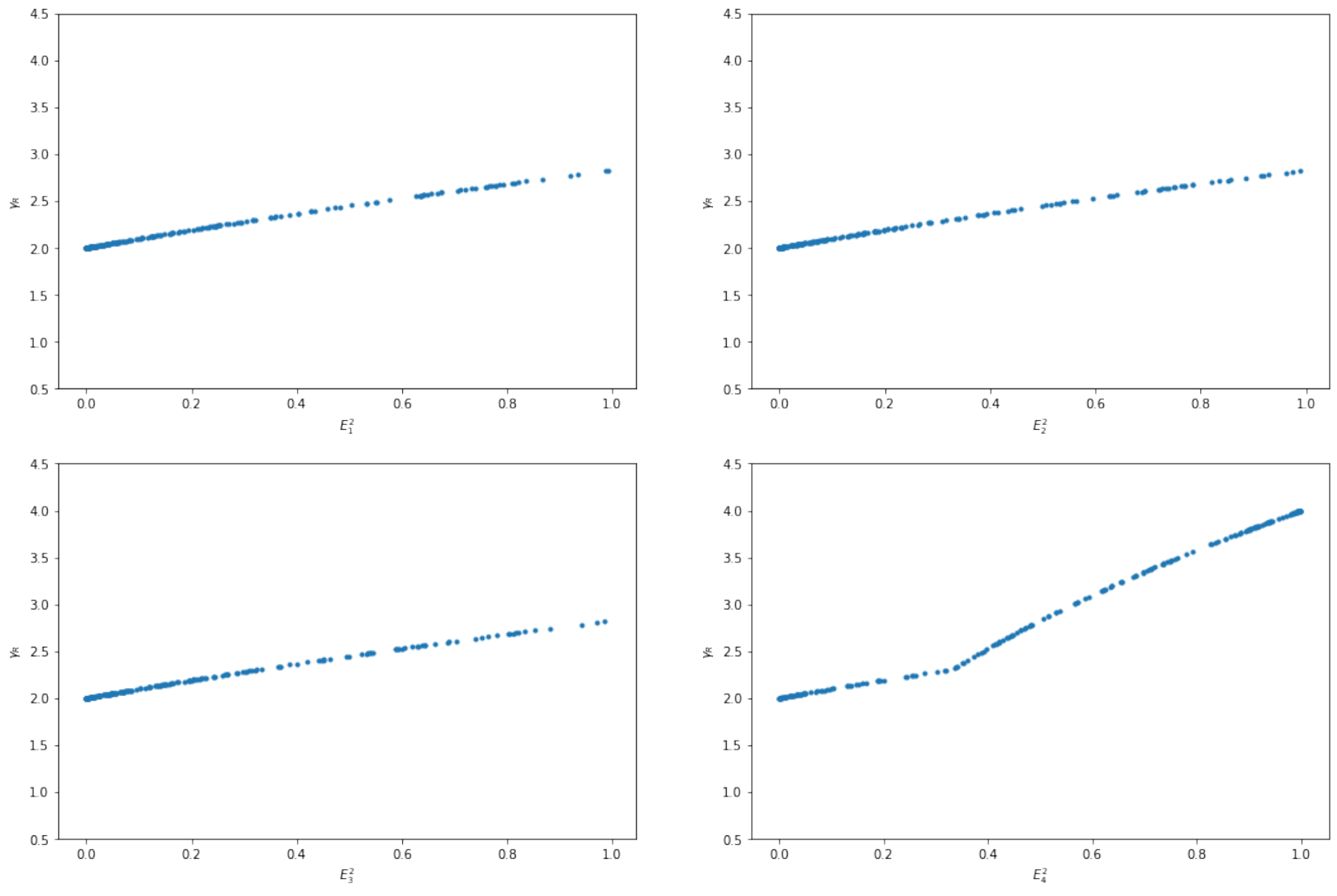}
\end{center}
\caption{We show that $\gamma_R$ restores the monotonically increasing behavior for $E_1^2$, $E_2^2$, $E_3^2$, and $E_4^2$. }
\label{gamma_r}
\end{figure}

\noindent 
Now we show the monotonic increasing result in Fig \ref{gamma_r}. 
The analytical solution also, in general, shows the monotonic increasing result for $-\alpha_1$ with a fixed $\gamma_2^{(j)}$ and $\theta^{(j)}$ in general. 
The LOCC showed that a general 3-qubit state has W-type and GHZ-type entanglement \cite{Dur:2000zz}. 
Therefore, we need to fix two parameters to indicate a choice of entanglement. 
The remaining parameter or total concurrence is to diagnose Quantum Entanglement. 
Therefore, Quantum Entanglement should be a source of $\gamma_R$ rather than the maximum violation $\gamma$. 

\section{Discussion and Conclusion}
\label{sec:4} 
\noindent
We showed that violating a constraint of correlations does not imply Quantum Entanglement. 
For our goal, we require a symmetric permutation of qubits. 
The 3-qubit operators are just a combination of four kinds of operators.  
Therefore, we can consider all cases without losing generality. 
Hence we then see how the maximum violation correlates with entanglement measures. 
We showed a loss of monotonically increasing. 
Here we only turn on one entanglement measure. 
In this case, the characterization of Quantum Entanglement does not have ambiguity. 
In other words, the monotone result holds when Quantum Entanglement is a necessary and sufficient condition for the violation. 
Our results showed that Quantum Entanglement is only a necessary condition. 
Hence we need to find an alternative measure to replace the violation. 
\\

\noindent    
The two largest eigenvalues of $R$-matrix \cite{Verstraete:2002} provides the maximum violation of Bell's inequality \cite{Bell:1964kc}. 
We generalized the $R$-matrix and provided an upper bound to maximum violation of Merlin's inequality. 
We then showed that the generalized $R$-matrix restores the loss behavior (monotonically increasing). 
Hence our result distinguishes the correlation of the $R$-matrix and maximum violation. 
The equivalence only holds in 2-qubit. 
The correlation of the generalized $R$-matrix is more proper to diagnose Quantum Entanglement than non-locality. 
We also rewrite the analytical solution ($\gamma_R$) in terms of five entanglement measures. 
This non-trivial fact shows that $\gamma_R$ contains all entanglement information.  
\\

\noindent 
When considering mixed states, not all entangled states lead to the violation of Bell's inequality. 
Therefore, entanglement (including mixed states) is necessary but not sufficient for the violation. 
Performing a partial trace operation on a 3-qubit state generates a 2-qubit mixed density matrix. 
We expect that the origin of ``Non-Locality$\neq$Quantum Entanglement'' may hide in a study of mixed states. 
One can use a partial trace operation to extend our analytical solution of generalized $R$-matrix to a 2-qubit mixed state. 
It should be interesting.    
\\

\noindent 
We proposed that the generalized $R$-matrix provides a proper diagnosis (Quantum Entanglement). 
Our result showed that the violation is not a possible diagnosis for pure state entanglement. 
Therefore, it also reflects the non-triviality of our proposal.    
The extension of $n$-qubits is simple in our proposal. 
Because a partial trace operation is unnecessary for measurement of $\gamma_R$, it simplifies an experimental study. 
Hence our proposal sheds light on exploring the mystery of many-body Quantum Entanglement. 
 
\section*{Acknowledgments}
\noindent 
We thank Xing Huang, Ling-Yan Hung, Masaki Tezuka, and Shanchao Zhang for their helpful discussion. 
Chen-Te Ma would like to thank Nan-Peng Ma for his encouragement.
\\

\noindent
Xingyu Guo acknowledges the Guangdong Major Project of Basic and Applied Basic Research No. 2020B0301030008 and NSFC Grant No.11905066.
Chen-Te Ma acknowledges the YST Program of the APCTP; 
Post-Doctoral International Exchange Program; 
China Postdoctoral Science Foundation, Postdoctoral General Funding: Second Class (Grant No. 2019M652926); 
Foreign Young Talents Program (Grant No. QN20200230017). 

\appendix 
\section{Numerical Results of Maximum Violation} 
\label{appa} 
\noindent 
We show all numerical results of maximum violation here without affecting the reading of the main context. 
The results show a loss of monotonic increase for the single entanglement measure case. 
\newpage
\begin{figure}
\begin{center}
\includegraphics[width=1.\textwidth]{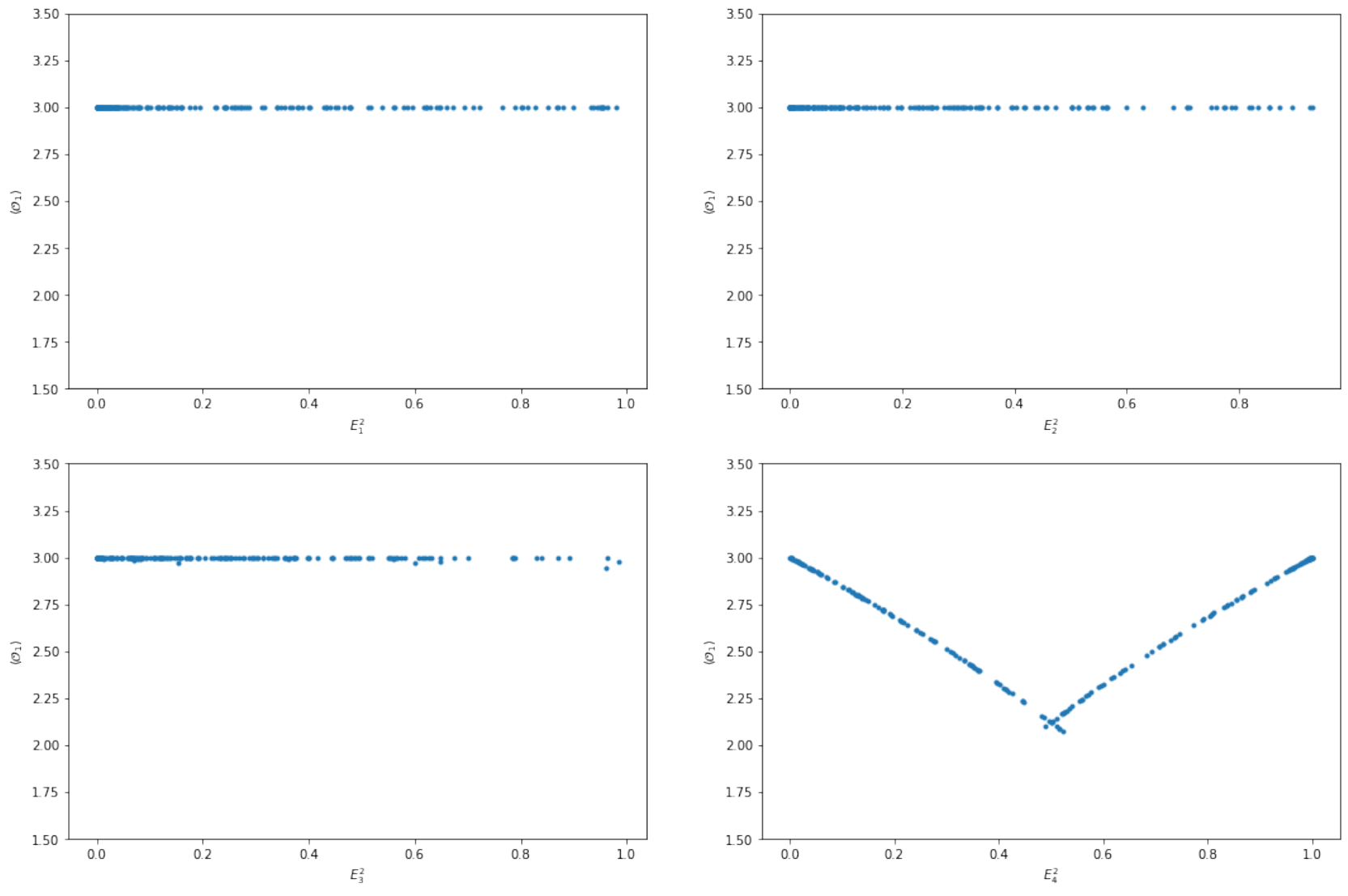}
\end{center}
\caption{We show $\langle{\cal O}_1\rangle$ for $E_1^2$, $E_2^2$, $E_3^2$, and $E_4^2$.}
\label{2-1}
\end{figure} 

\begin{figure}
\begin{center}
\includegraphics[width=1.\textwidth]{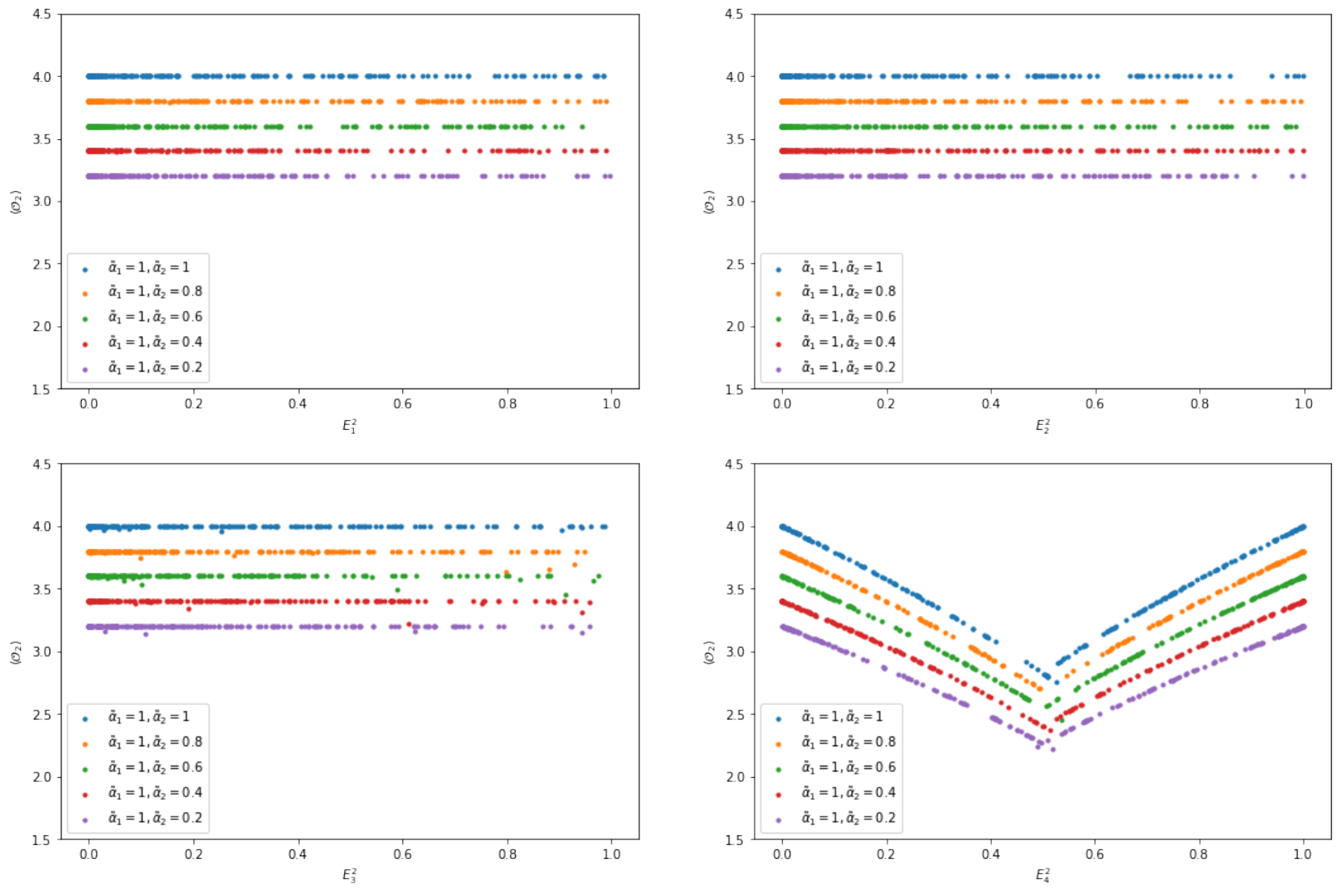}
\end{center}
\caption{We show $\langle{\cal O}_2\rangle$ for $E_1^2$, $E_2^2$, $E_3^2$, and $E_4^2$.}
\label{2-2}
\end{figure}
\clearpage

\begin{figure}
\begin{center}
\includegraphics[width=1.\textwidth]{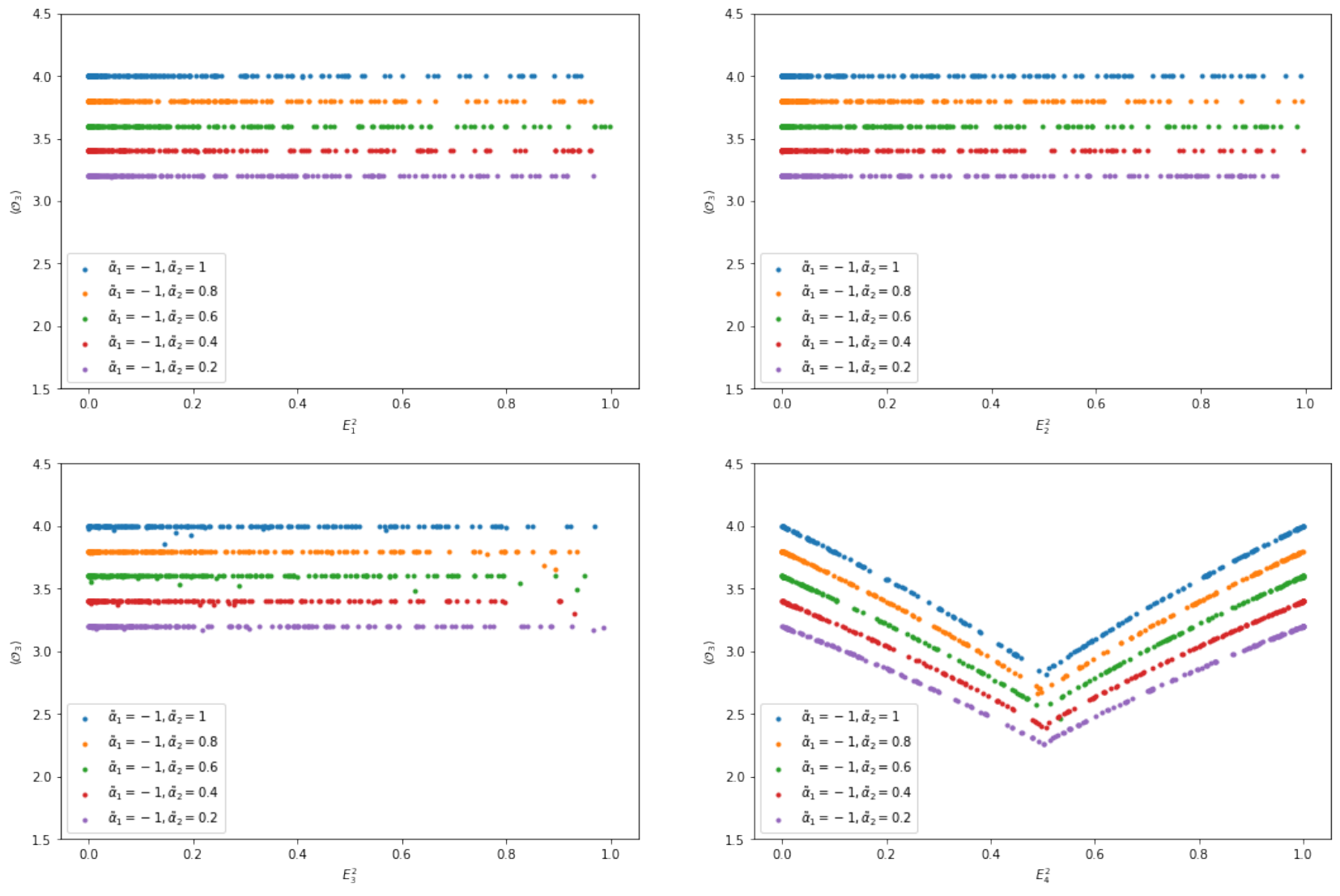}
\end{center}
\caption{We show $\langle{\cal O}_3\rangle$ for $E_1^2$, $E_2^2$, $E_3^2$, and $E_4^2$.}
\label{2-3}
\end{figure} 

\begin{figure}
\begin{center}
\includegraphics[width=1.\textwidth]{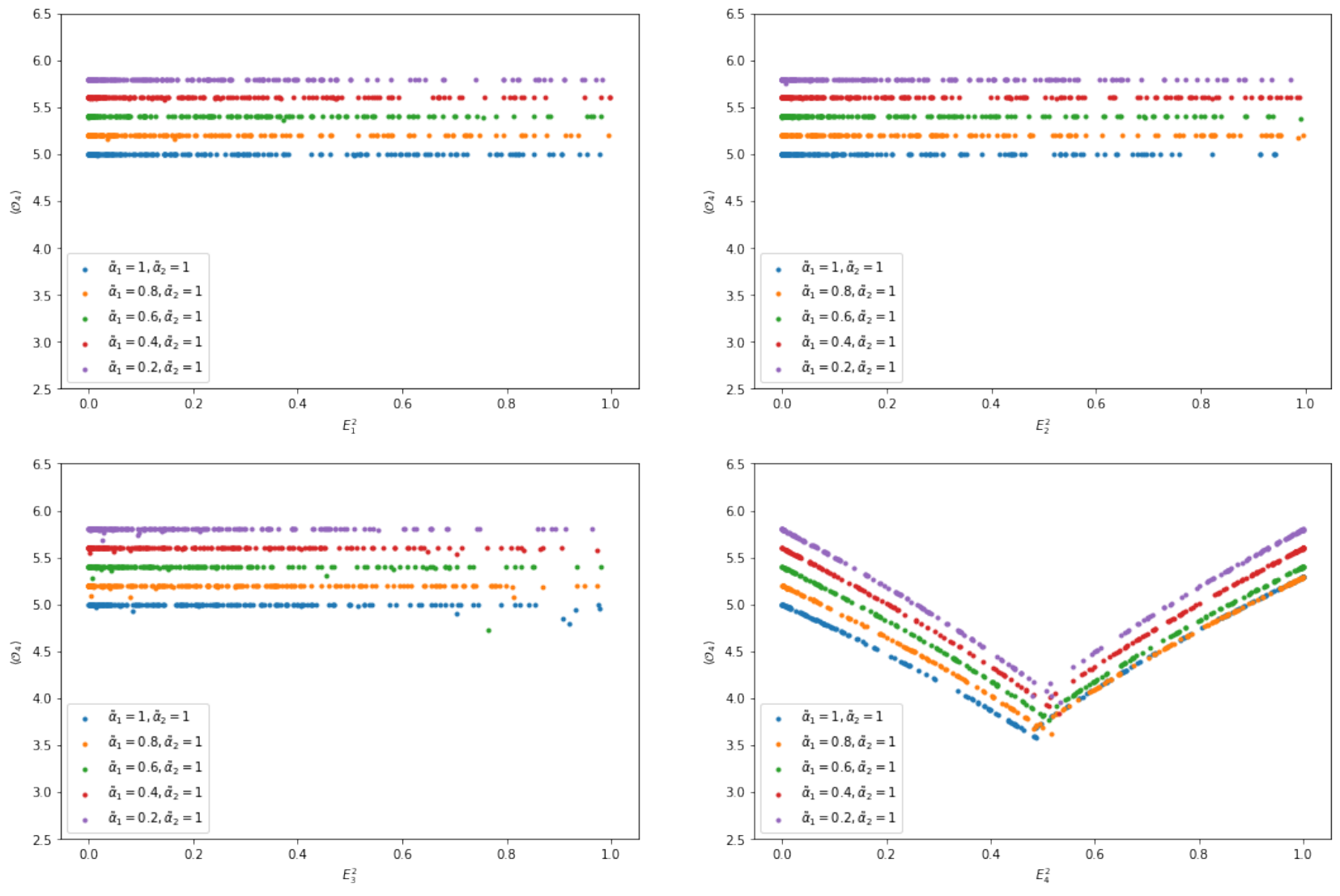}
\end{center}
\caption{We show $\langle{\cal O}_4\rangle$ for $E_1^2$, $E_2^2$, $E_3^2$, and $E_4^2$.}
\label{2-4}
\end{figure} 

\begin{figure}
\begin{center}
\includegraphics[width=1.\textwidth]{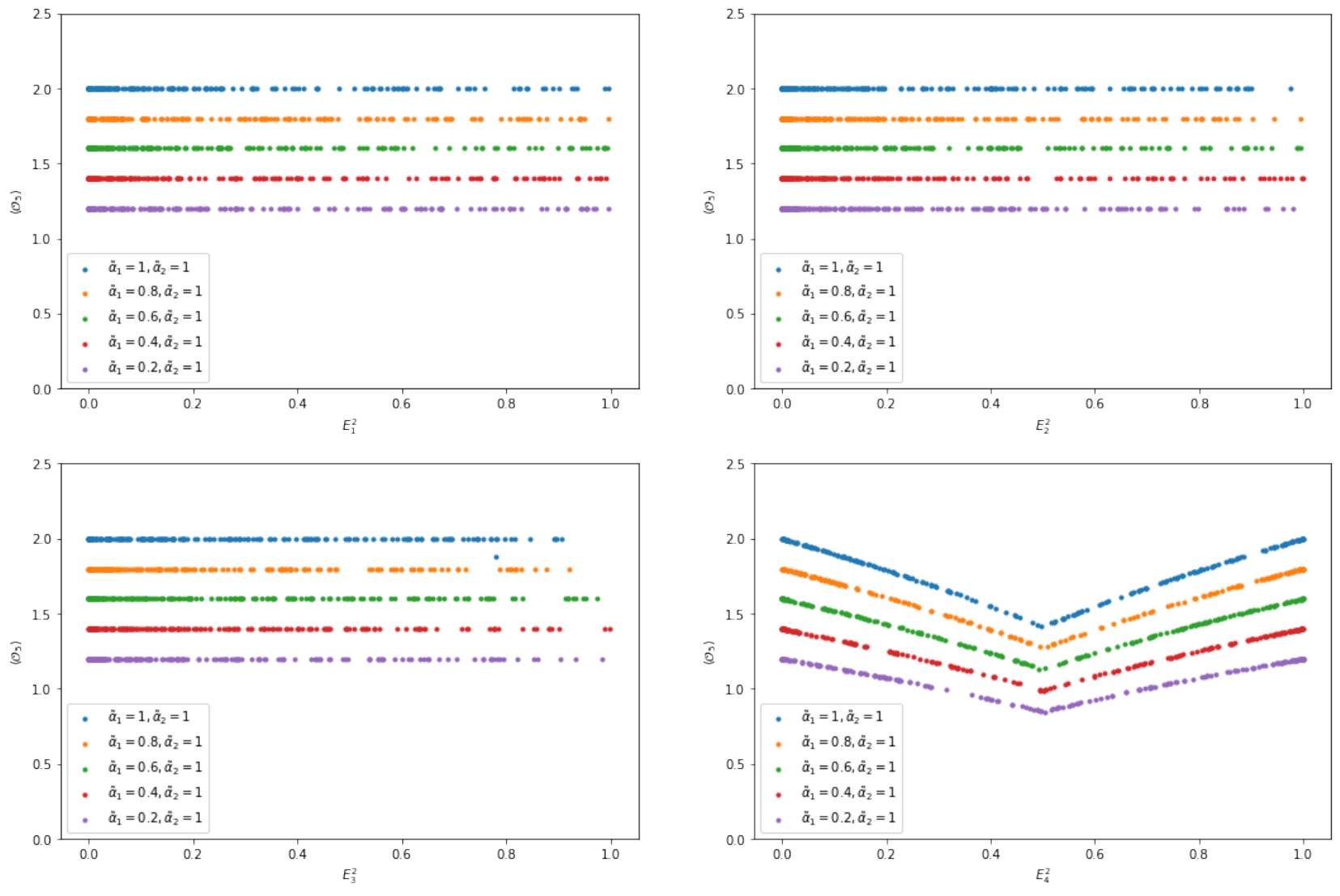}
\end{center}
\caption{We show $\langle{\cal O}_5\rangle$ for $E_1^2$, $E_2^2$, $E_3^2$, and $E_4^2$.}
\label{2-5}
\end{figure} 

\begin{figure}
\begin{center}
\includegraphics[width=1.\textwidth]{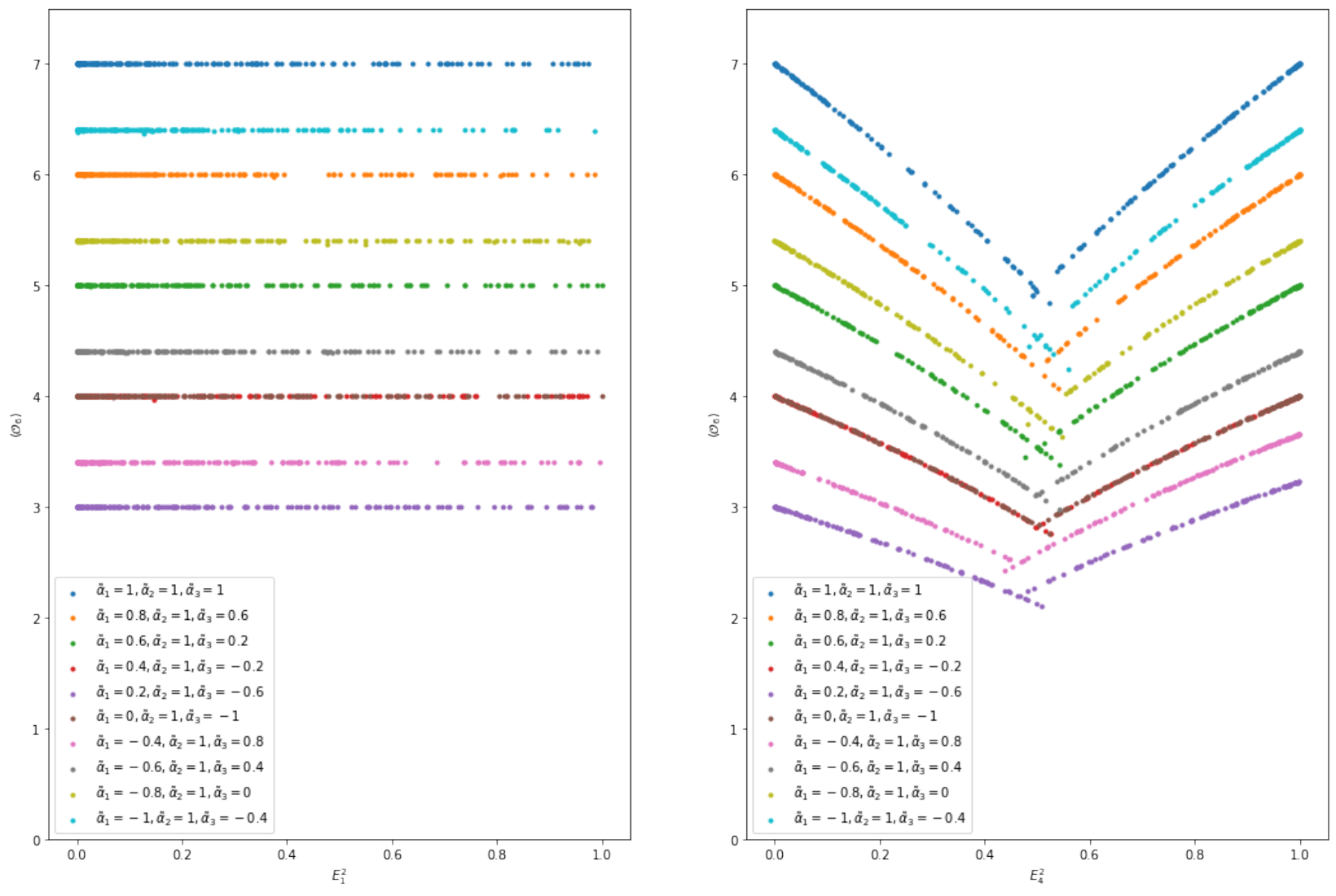}
\end{center}
\caption{We show $\langle{\cal O}_6\rangle$ for $E_1^2$ and $E_4^2$.}
\label{3-1}
\end{figure} 
\clearpage

\begin{figure}
\begin{center}
\includegraphics[width=1.\textwidth]{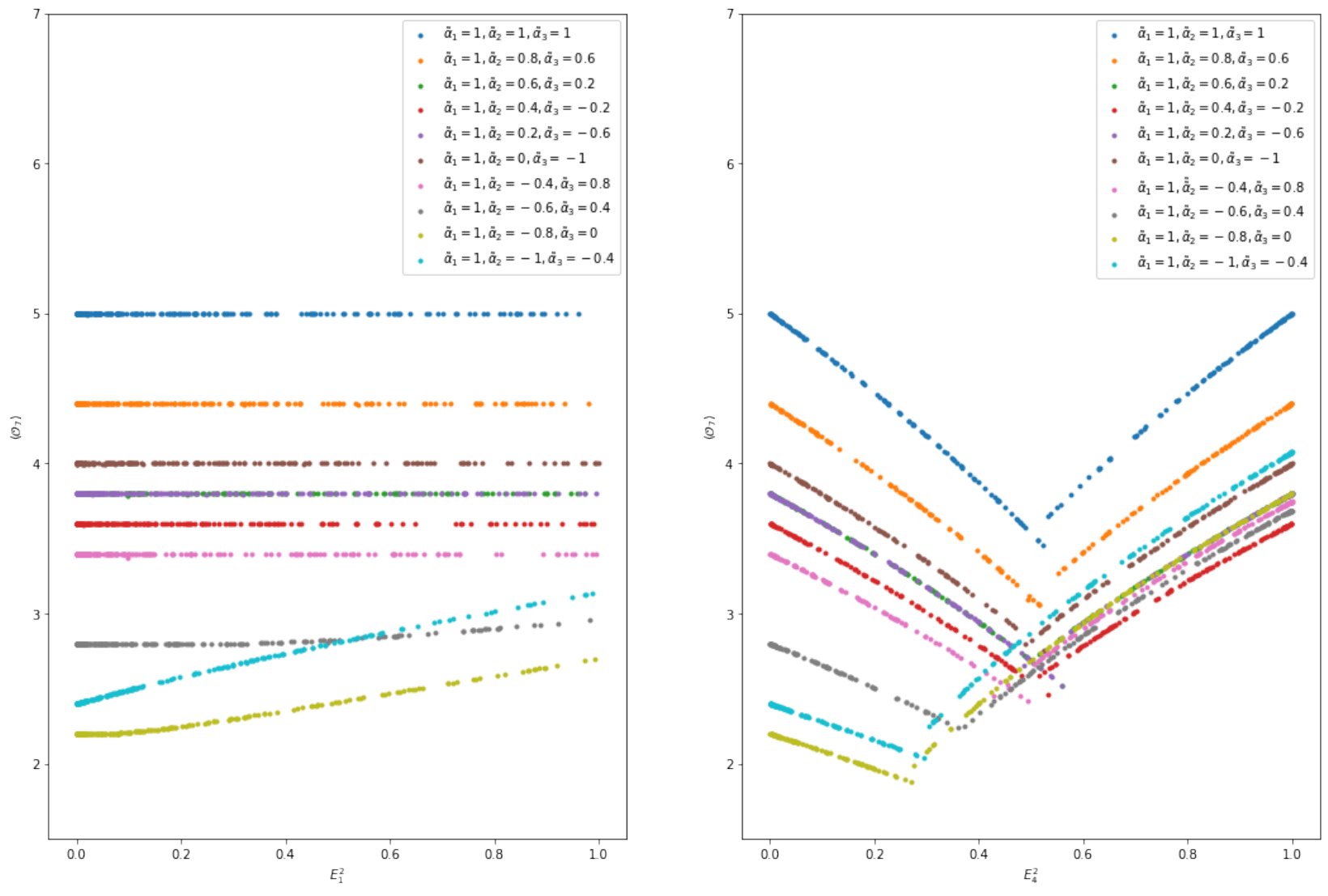}
\end{center}
\caption{We show $\langle{\cal O}_7\rangle$ for $E_1^2$ and $E_4^2$.}
\label{3-2}
\end{figure} 
\clearpage

\begin{figure}
\begin{center}
\includegraphics[width=1.\textwidth]{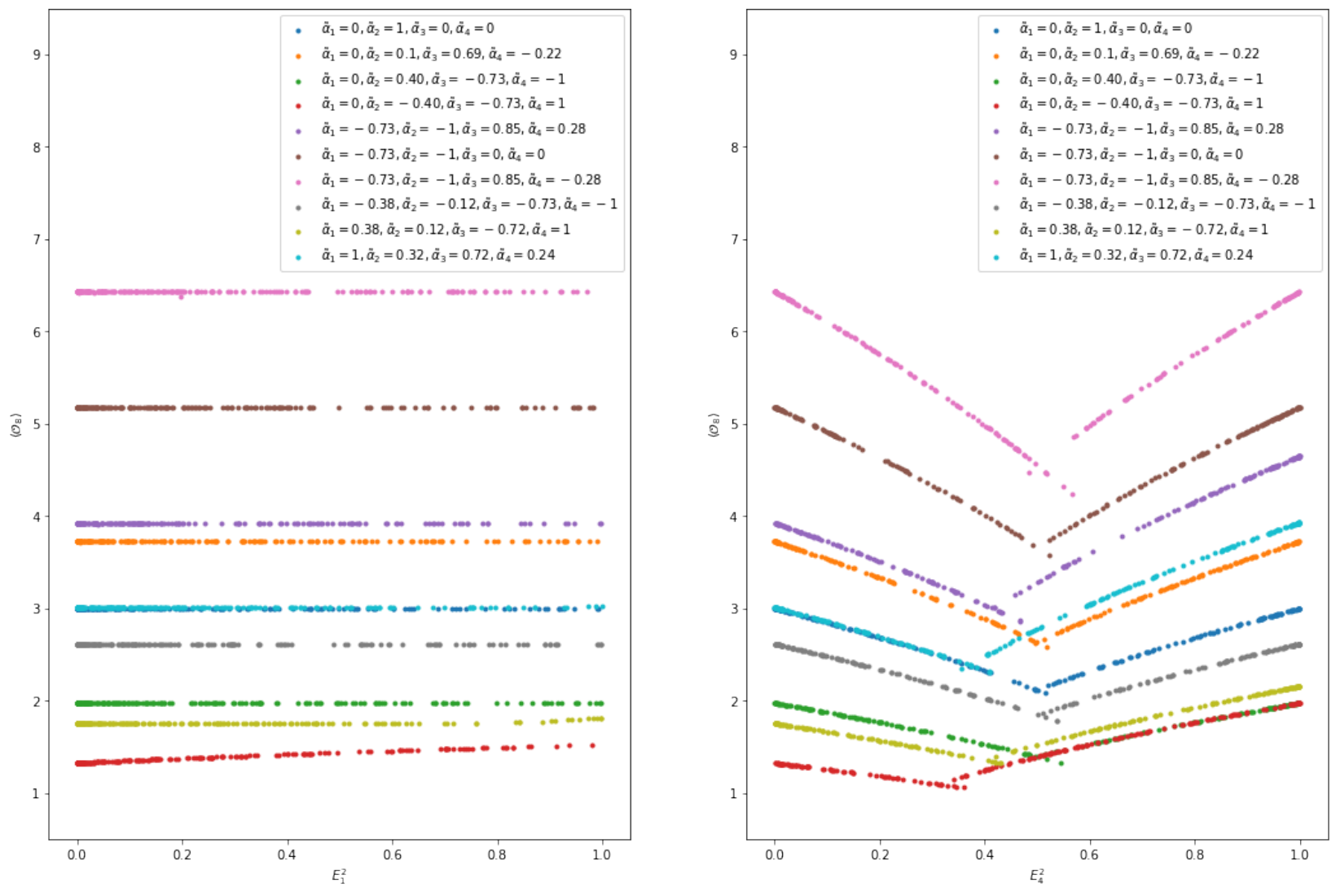}
\end{center}
\caption{We show $\langle{\cal O}_8\rangle$ for $E_1^2$ and $E_4^2$.}
\label{4}
\end{figure} 

\section{Calculation of $RR^T$} 
\label{appb}
\noindent
We first show the elements of $R_x$:
\bea
R_{xxx}&=&2\lambda_0\lambda_4;
\nn\\
R_{xxy}&=&0;
\nn\\
R_{xxz}&=&2\lambda_0\lambda_3;
\nn\\
R_{xyx}&=&0;
\nn\\
R_{xyy}&=&-2\lambda_0\lambda_4;
\nn\\
R_{xyz}&=&0;
\nn\\
R_{xzx}&=&2\lambda_0\lambda_2;
\nn\\
R_{xzy}&=&0;
\nn\\
R_{xzz}&=&2\lambda_0\lambda_1\cos(\phi).
\eea
We then show the elements of $R_y$:
\bea
R_{yxx}&=&0;
\nn\\
R_{yxy}&=&-2\lambda_0\lambda_4;
\nn\\
R_{yxz}&=&0;
\nn\\
R_{yyx}&=&-2\lambda_0\lambda_4;
\nn\\
R_{yyy}&=&0;
\nn\\
R_{yyz}&=&-2\lambda_0\lambda_3;
\nn\\
R_{yzx}&=&0;
\nn\\
R_{yzy}&=&-2\lambda_0\lambda_2;
\nn\\
R_{yzz}&=&2\lambda_0\lambda_1\sin(\phi).
\eea
We finally show the elements of $R_z$:
\bea
R_{zxx}&=&-2\lambda_1\lambda_4\cos(\phi)
-2\lambda_2\lambda_3;
\nn\\
R_{zxy}&=&2\lambda_1\lambda_4\sin(\phi);
\nn\\
R_{zxz}&=&-2\lambda_1\lambda_3\cos(\phi)
+2\lambda_2\lambda_4;
\nn\\
R_{zyx}&=&2\lambda_1\lambda_4\sin(\phi);
\nn\\
R_{zyy}&=&2\lambda_1\lambda_4\cos(\phi)
-2\lambda_2\lambda_3;
\nn\\
R_{zyz}&=&2\lambda_1\lambda_3\sin(\phi);
\nn\\
R_{zzx}&=&-2\lambda_1\lambda_2\cos(\phi)
+2\lambda_3\lambda_4;
\nn\\
R_{zzy}&=&2\lambda_1\lambda_2\sin(\phi);
\nn\\
R_{zzz}&=&\lambda_0^2-\lambda_1^2+\lambda_2^2+\lambda_3^2-\lambda_4^2=1-2\lambda_1^2-2\lambda_4^2. 
\nn\\
\eea
\\

\noindent
Now we calculate 
\bea
(R^{(1)}R^{(1)T})_{jk}\equiv \sum_{J}R^{(1)}_{jJ}R^{(1)}_{kJ}.
\eea
The result is:
\bea
&&(R^{(1)}R^{(1)T})_{xx}
\nn\\
&=&
4\lambda_0^2(\lambda_2^2+\lambda_3^2+2\lambda_4^2)+4\lambda_0^2\lambda_1^2\cos^2(\phi);
\nn\\
&&(R^{(1)}R^{(1)T})_{xy}
\nn\\
&=&(R^{(1)}R^{(1)T})_{yx}
\nn\\
&=&
4\lambda_0^2\lambda_1^2\cos(\phi)\sin(\phi); 
\nn\\
&&(R^{(1)}R^{(1)T})_{xz}
\nn\\
&=&(R^{(1)}R^{(1)T})_{zx}
\nn\\
&=&
2\lambda_0\lambda_1(2\lambda_1^2+2\lambda_4^2-1)-8\lambda_0\lambda_2\lambda_3\lambda_4
\nn\\
&&
+4\lambda_0\lambda_1(\lambda_2^2+\lambda_3^2+2\lambda_4^2)\cos(\phi)
;
\nn\\
&&(R^{(1)}R^{(1)T})_{yy}
\nn\\
&=&
4\lambda_0^2(\lambda_2^2+\lambda_3^2+2\lambda_4^2)+4\lambda_0^2\lambda_1^2\sin^2(\phi)
;
\nn\\
&&(R^{(1)}R^{(1)T})_{yz}
\nn\\
&=&(R^{(1)}R^{(1)T})_{zy}
\nn\\
&=&
2\lambda_0\lambda_1\sin(\phi)(1-2\lambda_0^2+4\lambda_4^2)
;
\nn\\
&&(R^{(1)}R^{(1)T})_{zz}
\nn\\
&=&
(1-2\lambda_1^2-2\lambda_4^2)^2
+4\big(\lambda_3\lambda_4-\lambda_1\lambda_2\cos(\phi)\big)^2
\nn\\
&&
+4\big(\lambda_2\lambda_4-\lambda_1\lambda_3\cos(\phi)\big)^2
\nn\\
&&
+4\big(\lambda_2\lambda_3-\lambda_1\lambda_4\cos(\phi)\big)^2
+4\big(\lambda_2\lambda_3+\lambda_1\lambda_4\cos(\phi)\big)^2
\nn\\
&&
+4\lambda_1^2\lambda_2^2\sin^2(\phi)
+4\lambda_1^2\lambda_3^2\sin^2(\phi)
+8\lambda_1^2\lambda_4^2\sin^2(\phi).
\nn\\
\eea

\end{document}